\documentclass[11pt, a4paper]{article}

\usepackage{datetime}
\newdateformat{monthyeardate}{\monthname[\THEMONTH] \THEYEAR}

\usepackage[utf8]{inputenc}
\usepackage{afterpage}
\usepackage{float}
\usepackage{pdflscape}
\usepackage{multirow}
\usepackage{footmisc} 
\usepackage[usenames,dvipsnames]{xcolor} 

\usepackage{caption}
\usepackage{subcaption}
\newcommand\extrafootertext[1]{%
    \bgroup
    \renewcommand\thefootnote{\fnsymbol{footnote}}%
    \renewcommand\thempfootnote{\fnsymbol{mpfootnote}}%
    \footnotetext[0]{#1}%
    \egroup
}

\usepackage{caption}
\usepackage{subcaption}
\usepackage{lipsum}

\usepackage[a4paper, left=2.54cm, right=2.54cm, top=2.54cm, bottom=2.54cm]{geometry}

\usepackage{graphicx}
\usepackage{amsmath}
\usepackage{amsthm}
\usepackage{amssymb}
\usepackage{hyperref}

\definecolor{mycolor}{rgb}{0.55, 0.0, 0.0} 

\hypersetup{
    colorlinks=true,       
    linkcolor=mycolor,     
    citecolor=mycolor,     
    filecolor=mycolor,     
    urlcolor=mycolor,      
}

\usepackage{comment}
\usepackage{booktabs}
\usepackage{authblk}
\usepackage{natbib}
\usepackage{setspace}
\usepackage{comment}
\usepackage[english]{babel}

\newtheorem{proposition}{Proposition}

\AtBeginDocument{%
  \setlength{\abovedisplayskip}{5pt}
  \setlength{\abovedisplayshortskip}{5pt}
  \setlength{\belowdisplayskip}{5pt}
  \setlength{\belowdisplayshortskip}{5pt}
}

\begin{document}
\title{Belief Bias Identification}
\author{Pedro Gonzalez-Fernandez}
\date{October 2026}
\affil{Heidelberg University}

\maketitle

\thispagestyle{empty}

\extrafootertext{For their invaluable support, I would like to thank Elias Tsakas, Thomas Meissner and Matthias Wibral. For helpful comments, I thank six anonymous referees, Mohammed Abdellaoui, Miguel Ballester, James Bland, Duarte Gonçalves, En Hua Hu, Peiran Jiao, Emmanuel Kemel, Yucheng Liang, Diego Marino-Fages, Antonio Penta, Arno Riedl, David Rojo Arjona, Michael Thaler, Severine Toussaert, Leeat Yariv and seminar participants at IMEBESS in Riga, ASFEE in Grenoble, Behavioural Game Theory Workshop in Norwich, Cognitive Foundations of Decision Making in Rabat, PhD conference in Micro Theory and Behavioral Economics in Berlin, Foundations of Utility and Risk in Alicante, and Maastricht University. I acknowledge funding from the Maastricht Graduate School of Business and Economics (GSBE) for primary data collection No. G.23.5168.

Homepage: \href{https://pgonzalezfernandez.com}{pgonzalezfernandez.com}. E-mail: \href{mailto:pedro.gonzalez-fernandez@awi.uni-heidelberg.de}{\texttt{pedro.gonzalez-fernandez@awi.uni-heidelberg.de}}. }

\begin{abstract}

This paper proposes a unified theoretical model to identify and test a comprehensive set of probabilistic updating biases within a single framework. The model achieves separate identification by focusing on the updating of belief distributions, rather than point beliefs alone. Estimating the model in a laboratory experiment reveals significant individual heterogeneity, with the hot-hand fallacy emerging as the most common bias and a key driver of biased inference. Accounting for additional distortions reveals associations across bias families and reduces the apparent prevalence of overinference and preference-based biases. At the population level, most biases average out, but base-rate neglect remains present. 


\end{abstract} 

\vspace{0.5\baselineskip}


\vspace{0.3cm}


\newpage

\pagenumbering{arabic}

\section{Introduction}

Over the last decades, substantial effort has gone into identifying and categorizing belief-updating biases, both within Economics and Psychology \citep{edwards1968conservatism,tversky1974judgment,grether1980bayes}. Scholars have typically studied these biases separately. As a result, it has become difficult to tell whether a person's updating behavior should be attributed to one bias or another. Despite recent progress toward more comprehensive representations of biased updating \citep{benjamin2019errors,stango2023we,bordalo2023memory,bordalo2025people, jakobsen2026coarse}, there is still a lack of a unifying framework that can separate multiple potentially conflicting biases within the same empirical setting.

The basic problem is that different belief-updating biases may be observationally equivalent. Consider a person who appears to update ``too much'' after receiving new information. A traditional model may ask whether this reflects base-rate neglect, namely underweighting prior beliefs, or overinference, meaning placing too much weight on the new signal. But other biases may generate a similar updating pattern. If the person has preferences over outcomes, is she overreacting because she generally overinterprets information, or because she is optimistic or pessimistic about the state she prefers? Is she instead reacting strongly because the signal confirms her prior beliefs? Or does the same posterior reflect unusually concentrated beliefs, that is, overprecision? Once some biases are left outside the model, behavior may be attributed to the wrong one. In that sense, some biases may appear empirically important partly because other competing distortions are not being modeled.

Distinguishing such confounded biases matters for at least two reasons. First, different biases can lead to very different actions even when they produce similar posteriors. For example, a person exhibiting confirmation bias may keep voting for the same political party after receiving social-media information that aligns with her previous beliefs, while someone who simply neglects prior information might be more likely to change her mind if reminded of the actual base rate. Second, once these biases are separately identified, one can ask which distortions are more prevalent, which matter most for inference, and which ones tend to appear together within individuals.

A central challenge in addressing this problem is methodological. The way beliefs are usually measured may not be rich enough to identify a broad set of potentially confounded biases. Much of the literature relies on point beliefs to study biased updating \citep{benjamin2019errors}. While convenient, point beliefs offer limited flexibility to distinguish between different updating patterns. This paper proposes an alternative framework based on belief distributions.

The paper constructs a structural model that allows multiple conflicting belief-updating biases to be separately identified using prior and posterior belief distributions.\footnote{While \citet{jiao2020social} use belief distributions to incorporate several biases simultaneously, their framework is not well suited to probabilistic biases, relies on normally distributed beliefs, incorporates a more limited set of distortions, and is not taken to experimental data.} Belief distributions can be understood either as beliefs over multiple outcomes or as uncertainty over one's own beliefs when the underlying state space is binary, in a sense related to cognitive uncertainty \citep{enke2023cognitive}. In either interpretation, the key idea is that belief distributions provide additional relationships for each updating decision, and a sharper language for defining and distinguishing belief biases. These additional relationships can identify biases separately with fewer updating environments.\footnote{Appendix \ref{app:identification_gain} formalizes this identification gain relative to binary-state belief reports.}

The theoretical framework considers an agent who observes a sequence of Bernoulli trials. Prior beliefs are assumed to follow a beta distribution and, under Bayesian updating, posteriors remain beta distributed because the beta distribution is conjugate to the binomial distribution.\footnote{This updating environment is comparable to settings used in the ambiguity-aversion literature (see \citet{abdellaoui2025learning} for an example) and in work on cognitive uncertainty \citep{enke2023cognitive}.} To introduce biases, the model allows for distortions of the likelihood and the prior. These distortions generate non-Bayesian posterior beta distributions and make it possible to identify core biases such as over- and underinference \citep{khaw2021individual,ba2022over,augenblick2025overinference, fan2026inference} and base-rate neglect or overuse \citep{benjamin2019base,bucher2022dynamics,enke2023baserate}. The framework also accommodates asymmetries in reactions to good and bad news, preference-based biases \citep{eil2011good,zimmermann2020dynamics,mobius2022managing}, confirmation bias \citep{rabin1999first,charness2017confirmation,Liu2024modelconfirmation}, sequence-related biases such as the hot-hand and gambler's fallacies \citep{rabin2002inference,rabin2010gambler,prat-carrabin2024resource}, and precision distortions \citep{moore2015overprecision,augenblick2025assumptions} that affect the variance of posterior beliefs.

The model is estimated using data from a novel laboratory experiment. Participants solve a series of urn tasks in which they must guess the percentage of red balls in a selected urn drawn from a pool of 99 urns with varying red-blue compositions. They receive information through sequences of ball draws from the selected urn and report their belief distributions twice: once after an initial signal, and again after observing a second sequence of independently drawn signals from the same urn.

The results support the idea that richer models of biased updating reduce the apparent importance of some biases found in more parsimonious specifications. At the population level, overinference and base-rate neglect are both significant when the model allows only these distortions. Once the model incorporates a broader set of biases, however, the evidence for overinference disappears and base-rate neglect remains as the only distortion at the aggregate level. This aggregate bias distribution is also broadly consistent with the classic reduced-form belief-updating literature when the same environment is translated into a traditional binary-state specification.

At the same time, aggregate results mask substantial individual heterogeneity.\footnote{The importance of individual heterogeneity for biased updating has recently been emphasized by \citet{khaw2021individual} and \citet{alos2023part}. Their work supports the idea that average reports may look almost Bayesian or noisy, while individual-level data reveal systematic but heterogeneous deviations that partly cancel out in the aggregate.} The individual-level analysis is therefore more informative, and changes the picture of which distortions matter most for updating. As in the aggregate analysis, overinference becomes less prevalent when a broader set of distortions is taken into account. Comparing intermediate specifications reveals which competing biases account for these changes: overinference and preference-based biases are particularly sensitive to the inclusion of sequence effects. By contrast, the hot-hand fallacy and base-rate neglect remain comparatively stable when other bias families are introduced.

This stability is particularly relevant because, as opposed to the aggregate findings, the hot-hand fallacy emerges as the most prominent bias across individuals: it is the most commonly exhibited distortion and generates substantial departures from Bayesian posterior expected value and variance. The hot-hand fallacy is also jointly exhibited with biases from several other families, and its presence provides information about whether other distortions are present in an individual's updating behavior. Nevertheless, pessimism and overinference also account for important departures in terms of magnitude, while base-rate neglect remains particularly relevant for posterior variance. Taken together, the results reveal substantial sequence and preference effects that are largely obscured by the aggregate analysis, and show that allowing for sequence effects changes how other departures from Bayesian updating are classified.

The paper contributes to the literature in two main ways. First, it contributes to recent work that seeks to understand the links between different updating distortions. For example, \citet{heger2018we} and \citet{gneezy2023can} study how wishful thinking can affect overconfidence; \citet{charness2017confirmation} and \citet{Liu2024modelconfirmation,Liu2023dualelicitation} distinguish behavior stemming from motivated beliefs and unmotivated confirmation bias; \citet{aydogan2025much} build on \citet{rabin1999first} to separate conservatism from confirmation bias; and \citet{chopra2024demand} study the trade-off between accuracy concerns and belief-confirmation motives in political information processing. These are important steps toward understanding how biases interact. More generally, however, in the absence of a framework that accommodates a broad range of distortions simultaneously, empirical links between biases may themselves be confounded. Bias $A$ may appear related to bias $B$ only because a third bias $C$ is omitted. This approach is also related in spirit to \citet{chapman2023econographics}, who study the joint correlation structure of a broad set of behavioral regularities to provide an empirical basis for more comprehensive theories of decision-making. The present paper applies a similar perspective within belief updating, while additionally addressing the problem that different updating distortions may be empirically confounded when modeled separately.

Second, the paper contributes to research that seeks belief-based explanations for behavioral phenomena in settings where several competing distortions are plausible. Political polarization, for instance, has been explained from the perspective of overconfidence \citep{ortoleva2015overconfidence}, which is closely related to overprecision, and also from the perspective of confirmation bias \citep{taber2006motivated,del2017modeling}, and motivated reasoning \citep{kahan2013ideology}. In finance, the disposition effect has been linked to motivated beliefs \citep{heinke2023belief}, the gambler's fallacy \citep{jiao2017belief}, and general underinference \citep{pitkajarvi2022slow}. Other examples include work linking confirmation bias to stylized facts in financial markets \citep{pouget2017mind}, or confidence distortions to poor investment performance \citep{ahmad2020overconfidence} and biased memory \citep{huffman2022persistent}. A framework that distinguishes conflicting biases more cleanly can therefore help identify which distortions are actually driving particular behavioral regularities, and whether some candidate explanations are partly artifacts of less complete models.

The remainder of the paper is organized as follows. Section \ref{sec.Theory} introduces the theoretical framework and progressively incorporates different biases into the model. Section \ref{sec.design} presents the experimental design and the belief-measurement tool. Section \ref{sec.methodology} brings the theoretical and experimental components together and compares two regression models that differ in the number of biases they incorporate. Section \ref{sec: Results} discusses the results, and Section \ref{sec. conclusion} concludes.

\section{A model of multiple belief biases}
\label{sec.Theory}

\subsection{Theoretical framework}
\label{sec.framework}

Let an agent observe a signal $S = \{s_1, ..., s_n\}$ consisting of the realization of $n$ independent and identically distributed Bernoulli trials. Let $p$ denote the probability of success of each trial ($s_i=1$) and $1-p$ denote the probability of each failure ($s_i=0$). For such a data-generating process (DGP), the likelihood function is the probability mass function of a binomial distribution with parameters $(n,p)$:
\begin{equation}
\label{eq.like}
    L(p|s_1, \dots, s_n) = \binom{n}{k} p^{k} (1-p)^{n - k}
\end{equation}
where $k = {\sum_{i=1}^{n}s_i}$ and $(n - k)$ are the number of successes and failures in the DGP respectively.

Let $\Omega = (0,1)$ be the set of possible values that $p$ may take,\footnote{The state-space set $\Omega$ can have two different interpretations. One could either consider an agent who is not uncertain about her own beliefs, or one who is indeed uncertain about her own beliefs (à la cognitive uncertainty \citep{enke2023cognitive}). In the former case, the set $\Omega$ would specify the state-space of an agent who forms beliefs over every possible realization of the objective parameter $p \in (0,1)$. In the latter case, the agent would form beliefs over a binary state-space (whether the signal $s_i$ takes value $1$ or $0$), and the set $\Omega$ would, in this case, represent the subjective state-space ($p$), which describes the set of possible states where the agent expresses uncertainty about her own beliefs. The model is agnostic to either one of these interpretations.} and let a prior belief $\pi(p)$ be beta distributed with prior parameters $(a_0,b_0)$. Namely,
\begin{equation}
\label{eq.prior}
    \pi(p|a_0, b_0)=
       \frac{1}{B(a_0, b_0)}p^{a_0 -1}(1-p)^{b_0 -1}
\end{equation}
where $a_0$, $b_0$ $ > 0$, and $B(.)$ is the beta function.

Given a prior $\pi(p)$ and a likelihood $L(p|s_1 ... s_n)$ the agent forms a posterior. As the beta distribution is a conjugate prior of the binomial distribution, a Bayesian agent updates her beliefs such that her posterior distribution of $p$, $\pi(p| a_n, b_n)$,  is also beta distributed with parameters $a_n, b_n$. This means that:\footnote{See Appendix \ref{sec.Conjugate analyses} for a proof; as well as a proof to derive eq.\eqref{eq.non-bayes S_a} and \eqref{eq.non-bayes S_b}, which suffice to derive all other equations of posterior parameters in section \ref{sec.Theory}.} 
\begin{equation}
\label{eq.bayes prmt}
    a_n = k + a_0 \qquad b_n = n-k + b_0  
\end{equation}
In order to incorporate updating biases, let a non-Bayesian agent use a distorted likelihood and prior when she updates.\footnote{These distortions are often interpreted \citep{benjamin2019errors} as part of an ``as-if model.'' Accordingly, the bias parameters capture systematic deviations from Bayesian updating, rather than the underlying psychological mechanisms generating those deviations. In particular, the model does not assume that biased agents literally apply Bayes' Theorem with distorted likelihood and prior functions; these distortions instead provide an equivalent representation of behavior in which agents place too little or too much weight on information signals (or prior beliefs) when updating. Related models instead derive distorted inference from misspecified beliefs about the sampling process; see, for example, \citet{benjamin2016model}.} These distortions can be expressed as exponential deviations of the  likelihood and prior. Namely, $\Tilde{L}(p|s_1, \dots, s_n) = (L(p|s_1, \dots, s_n))^\gamma$ would represent the distorted likelihood and $\Tilde{\pi}(p) = (\pi(p))^\delta$ would represent a distorted prior of the non-Bayesian agent ($\gamma, \delta \in \mathbb R$). The parameters $\gamma \neq 1$ and $\delta \neq 1$ indicate deviations from Bayesian updating due to distortions of the likelihood and prior respectively. With these modified functions such a non-Bayesian agent has a posterior beta distribution with parameters $\Tilde{a}_n,\Tilde{b}_n$ such that:
\begin{equation}
\label{eq.non-bayes S_a}
    \Tilde{a}_n = \gamma k + \delta (a_0 - 1) + 1
\end{equation}
\begin{equation}
\label{eq.non-bayes S_b}
    \Tilde{b}_n = \gamma (n -k) + \delta (b_0 - 1) + 1
\end{equation}
In particular $\gamma > 1$ indicates ``overinference" while $\gamma \in [0,1)$ shows ``underinference"; that is, believing that the information signal $S$ is more/less informative than it actually is. A parameter of $\gamma <0$ reflects updating against the information signal. Similarly, parameter $\delta > 1$ indicates ``base-rate overuse" and $\delta < 1$, ``base-rate neglect"; which implies that the prior is more/less informative than a Bayesian agent perceives it to be. This bias structure conceptually resembles Grether-regressions,\footnote{That is, binary-state regressions where inference biases and base-rate biases are identified.} as the same type of biases can be identified. The next section shows how equations \eqref{eq.non-bayes S_a} and \eqref{eq.non-bayes S_b} can be extended to incorporate incrementally more biases.     

\subsection{Introducing multiple belief biases}
\label{sec.model multiple}

The main advantage of this representation is that the two posterior parameters provide separate updating equations (i.e. for successes and failures) within each updating decision. Appendix \ref{app:identification_gain} shows formally how this approach reduces the number of belief-updating decisions required to identify multiple updating distortions relative to a scalar binary-state representation.

\subsubsection{Asymmetries between successes and failures \& Preference-based biases} 

Equations \eqref{eq.non-bayes S_a} and \eqref{eq.non-bayes S_b} assume that deviations from the Bayesian agent in the likelihood are symmetric for successes and failures. That is, the agent over or under-reacts to ``positive information" the same way she over or under-reacts to ``negative" information. For a non-Bayesian agent this need not be the case. Consider instead that the agent weights successes and failures of the data generating process differently. Then her likelihood function would be  $\Tilde{L}(p|s_1, \dots, s_n) \propto p ^{\alpha k} (1-p)^{\beta (n - k)}$.  In turn, her posterior would be beta distributed with shape parameters $\Tilde{a}_n,\Tilde{b}_n$ such that:
\begin{equation}
\label{eq.non-bayes S_alpha}
    \Tilde{a}_n = \alpha k + \delta (a_0 - 1) + 1
\end{equation}
\begin{equation}
\label{eq.non-bayes S_beta}
    \Tilde{b}_n = \beta (n -k) + \delta (b_0 - 1) + 1
\end{equation}
where $\alpha \neq \beta$ indicates asymmetric reactions to successes and failures in the DGP. Equations \eqref{eq.non-bayes S_alpha} and \eqref{eq.non-bayes S_beta} are especially interesting if the agent has preferences over the state-space. Suppose this is the case, and suppose further that preferences are expressed by a utility function, which is continuous and monotonically increasing over $p$. Then, every success $k$ would be informative of a higher realization of $p \in (0,1)$, i.e. a preferred state. Conversely, every failure $(n-k)$ is informative of a lower realization of $p$. Therefore, successes and failures can be interpreted as pieces of good and bad news respectively. This means that $\alpha>1$ (overreacting to positive information) or $\beta <1$ (underreacting, or updating against the signal when confronted with negative information) can be interpreted as \textit{optimism} bias; while $\alpha <1$ or $\beta>1$ would indicate \textit{pessimism}. Furthermore, the model also captures \textit{motivated asymmetries}: $\alpha > \beta$ represents the good news effect, while $\alpha < \beta$ implies there is a bad news effect.\footnote{Asymmetries between successes and failures can also be present at the level of the prior. These distinctions are introduced in section \ref{sec.methodology}, but left out of the main theoretical discussion. This is without loss of generality as the resulting equations are identical to equations \eqref{eq.non-bayes Conf_alpha} and \eqref{eq.non-bayes Conf_beta } except for having separate $\rho$ and $\delta$ parameters for successes and failures (see Section \ref{sec.Meth complete})} 

\subsubsection{Confirmation Bias}
\label{sec. Confirmation}
\begin{figure}[!htbp]
    \centering
    \includegraphics[width=10cm]{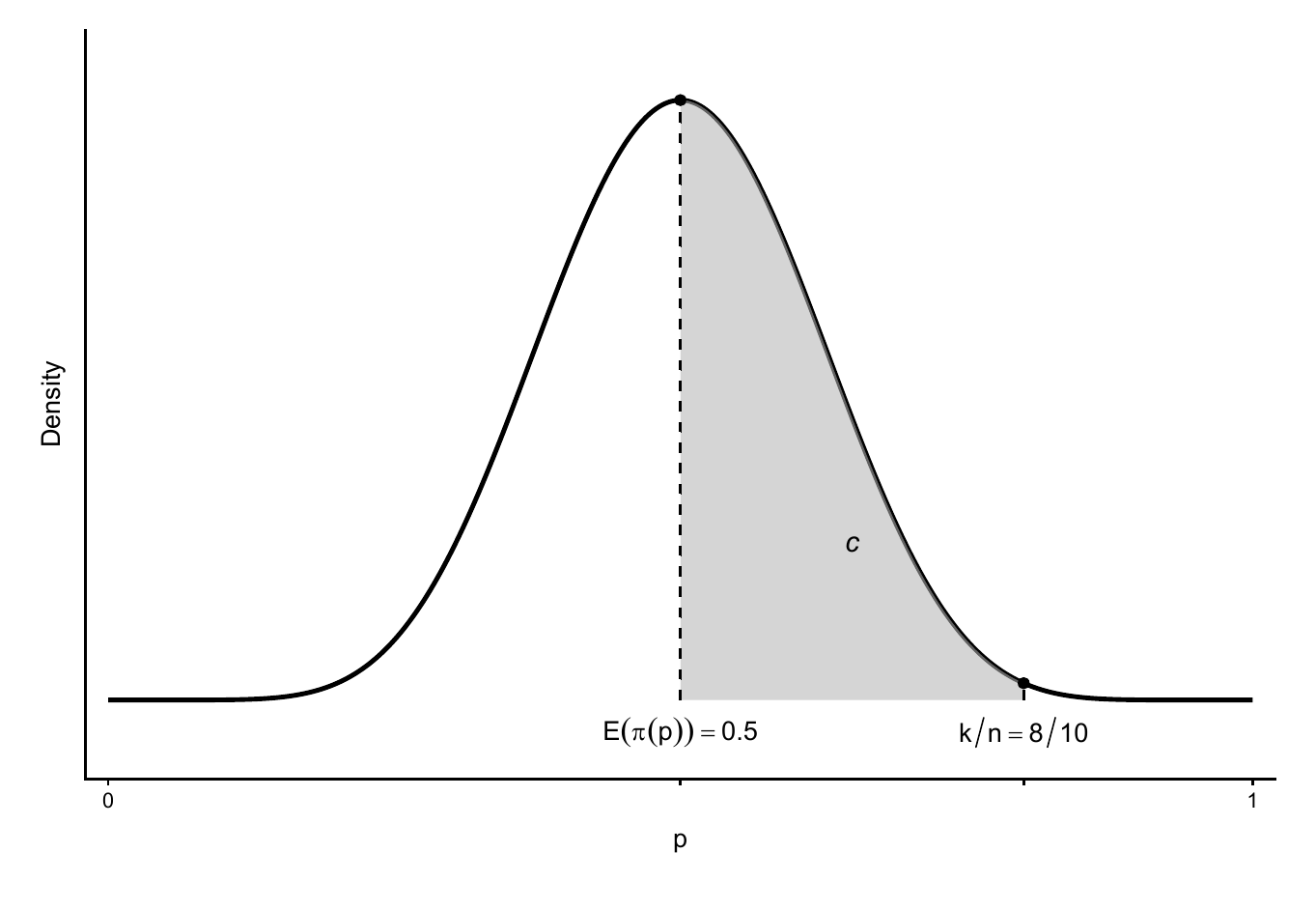}
    \caption{Example of confirmation measure for an updating problem with $a_0=b_0=9$, prior mean $E(\pi(p))=0.5$ and information signal of 8 successes and 2 failures ($k/n=8/10$)}
    \label{fig:confirmation}
\end{figure}

Confirmation bias is modeled as a positive correlation between overreaction to information signal $S$, and how confirming the signal is. The degree of confirmation of the signal is unrelated to the agent holding preferences\footnote{A discussion about the distinction between motivated and unmotivated confirmation bias can be found in \citet{Liu2023dualelicitation}.} over the state-space. In particular, the degree of confirmation $c$ is expressed as the area, in the density function of the prior, comprised between the expected value of the prior $E(\pi(p))$ and the mean of the information signal $k/n$. Formally:
\begin{equation}
\label{eq.Confirmation variable}
    c= \bigg|\int_{(k \pm \varepsilon)/n}^{E(\pi(p))} \pi(p|a_0,b_0) \,dp \bigg|
\end{equation}
Equation \eqref{eq.Confirmation variable} specifies a relative measure of confirmation\footnote{In equation \eqref{eq.Confirmation variable}, the signal mean $k/n$ is replaced by $(k \pm \varepsilon )/n$. This is because $p$ is not strictly defined at $p=0$ and $p=1$. Thus, one must take $(k + \varepsilon )/n$ as the lower integration limit if $k=0$; and $(k - \varepsilon )/n$ if $k=n$.} (see Figure \ref{fig:confirmation} for an example). The \textit{higher} the value of $c$, the \textit{less confirming} a signal will be. To incorporate over or under reaction to confirmation a separate term $\rho c$ is added to the distorted likelihood function. Equations \eqref{eq.non-bayes S_alpha} and \eqref{eq.non-bayes S_beta} are modified as follows:
\begin{equation}
\label{eq.non-bayes Conf_alpha}
    \Tilde{a}_n = \alpha k + \rho  c + \delta (a_0 - 1) + 1
\end{equation}
\begin{equation}
\label{eq.non-bayes Conf_beta }
    \Tilde{b}_n = \beta (n -k) + \rho  c + \delta (b_0 - 1) + 1
\end{equation}
where $\rho<0$ indicates confirmation bias (as $c$ becomes smaller the perception of the number of successes or failures grows); and $\rho>0$ indicates disconfirmation bias. 

\subsubsection{Sequence-related biases and unrelated inference biases}
\label{sec.biases seq and pref}

Overreaction or underreaction to information signals can also be driven by reasons other than preferences over the state-space. I distinguish between \textit{sequence-related biases} (e.g., the hot-hand fallacy and the gambler's fallacy) and \textit{baseline inference}. That is, overreaction or underreaction to information signals when preferences over the state-space are not at stake. 

\vspace{2mm}

\textbf{Sequence-related biases:} Consider a partition of the signal space $S$. Namely, $S_1 = \{s_1 \dots s_m\}$ and $ S_2 = \{s_{m+1} \dots s_n\}$ where $n>m$. Suppose all of the $s_i \in S_2$ are either successes (i.e. $\sum_{m+1}^{n} s_i= n-m$, $s_i\in S$) or failures (i.e. $\sum_{m+1}^{n} s_i= 0$, $s_i\in S$). In this context, the hot-hand fallacy is defined as inferring too much from information signals after observing the last $n-m$ consecutive successes (i.e. $\alpha > 1$) or failures (i.e. $\beta > 1$). Conversely, gambler's fallacy is defined as underinferring after observing the last $n-m$ consecutive successes (i.e. $\alpha < 1$) or failures (i.e. $\beta < 1$). Thus, sequence-related biases are modeled as a context-dependent component of inference distortions, rather than as a separate channel.

\vspace{2mm}

\textbf{Inference without preferences (baseline inference):}
Consider an almost equivalent state-space set, $\Omega_{NP} = (0,1)$ over which the agent forms beliefs but where, contrary to $\Omega$, the agent does not hold any preference. So any element $\hat{p}$ of the set $\Omega_{NP}$ will have the characteristic that $u(\hat{p})=0$. Because this is the only difference between sets, only $\alpha$ and $\beta$ coefficients in equations \eqref{eq.non-bayes Conf_alpha} and \eqref{eq.non-bayes Conf_beta } differ between forming beliefs over $\Omega$ or $\Omega_{NP}$. Therefore, when the agent forms beliefs over $\Omega_{NP}$, $\alpha$ and $\beta$ coefficients can be interpreted as over(under) inference for successes and failures, independently of preferences over the state-space.

\vspace{2mm}

The biases in this subsection are only separably identified when the agent faces different updating environments (i.e. multiple belief-elicitation decisions) that generate variation both in whether preferences are at stake (tasks in $\Omega$ versus $\Omega_{NP}$) and in whether the observed signal contains a streak.

\subsubsection{Under and overprecision}
\label{sec.precision}

Precision biases (overprecision and underprecision) are those which are strictly related to the variance of the agent's posterior distribution in relation to the Bayesian variance.\footnote{As an alternative way to capture deviations in the second moment of the distribution, one could also use the concentration parameter of the Bayesian posterior ($\kappa_n=a_n+b_n$) to define under and overprecision. This alternative definition has the advantage of isolating precision biases within equations \eqref{eq.non-bayes Conf_alpha} and \eqref{eq.non-bayes Conf_beta }. (i.e. Underprecision: $\alpha+\beta<2$, and Overprecision: $\alpha+\beta>2$). However, it imposes harsh restrictions on $\alpha$ and $\beta$ parameters, making precision biases harder to detect.} Underprecision implies that the agent's overall distribution of posterior beliefs is more dispersed than that of a Bayesian agent. Conversely, overprecision yields a posterior distribution which is less dispersed over the values of $p$ than the distribution a Bayesian agent would have. That is:
\begin{equation}
\label{eq.Var}
        \Tilde{Var_n} = \nu \times Var_n
\end{equation}
where $\Tilde{Var_n}$ is the variance of the agent's posterior beta distribution, and $Var_n$ is the Bayesian variance. In equation \eqref{eq.Var}, $\nu<1$ indicates overprecision, while $\nu>1$ indicates underprecision.

\paragraph{Summary.}
For ease of reference, Table~\ref{tab:summary biases} in Section~\ref{sec.Meth complete} summarizes the mapping from parameter restrictions to the bias labels used throughout the paper.

\section{Experimental Design}
\label{sec.design}

The experiment provides a setting in which the belief biases described in the theoretical framework of Section \ref{sec.Theory} can be identified at both the individual and population levels. It was conducted at the Behavioral and Experimental Economics Laboratory (BEELab) at Maastricht University. A total of 88 participants were recruited, and each completed 30 belief-elicitation tasks. In each task, participants reported their belief distributions twice, after observing two consecutive information signals. The average payment per participant was 15.9 euros. The experiment was pre-registered in October 2023.\footnote{See the preregistration at \hyperlink{https://aspredicted.org/285v-b8cg.pdf}{https://aspredicted.org/285v-b8cg.pdf}} The full instructions of the experiment can be found in Appendix \ref{sec. instructions}. 

\subsection{The belief-elicitation task}
\label{sec. task}

\begin{figure}[ht]
    \centering
    \includegraphics[width=1\linewidth]{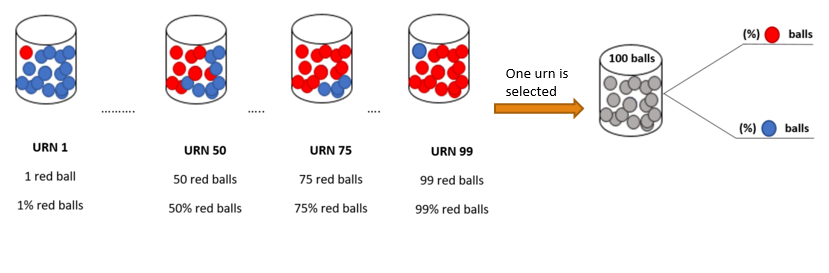}
    \caption{Urn selection}
    \label{fig:URN}
\end{figure}

In each of the 30 tasks, participants observed a pool of 99 urns, each containing 100 balls. This provides a discrete implementation of the continuous state space introduced in Section \ref{sec.Theory}. Each urn contained a different composition of red and blue balls. Thus, Urn 1 contained 1 red ball and 99 blue balls, Urn 2 contained 2 red balls and 98 blue balls, and so on, up to Urn 99, which contained 99 red balls and 1 blue ball. One of these urns was then selected at random from the pool, but its content was not revealed to participants. Their task was to guess the percentage of red balls in the selected urn (see Figure \ref{fig:URN}). To do so, participants received information signals by observing two sequences of balls drawn with replacement. In the first sequence, either one, two, or three balls were drawn at random from the selected urn.

After observing this first sequence, participants were asked to report their belief distribution\footnote{See Section \ref{sec.eliciting} for details about the elicitation of beta distributions.} (Figure \ref{fig:Prior example}) about the percentage of red balls in the selected urn. This first report is treated as the prior for the second updating stage, so that the analysis captures incremental updating from subjects' own previously reported beliefs.\footnote{This treatment is consistent with the ``acceptive'' representation of sequential updating discussed by \citet{benjamin2019errors}, in which the posterior after one signal becomes the prior for the subsequent update; see also \citet{raymond2024optimal}. Using instead the exogenously uniform initial prior would eliminate variation in prior beliefs and hence the variation needed to identify prior-weighting distortions. Consistent with this interpretation, the participant's first report was also displayed as the default distribution for the second elicitation.} Once this prior distribution was elicited, a second sequence of draws from the same selected urn was shown to participants. In this case, either three, five, or seven balls were drawn at random from the urn. After observing this second information signal, participants were asked to report their posterior belief distribution\footnote{Both in the prior and posterior reports, participants needed to update from a given default beta distribution. Before eliciting their prior, the default beta distribution shown was uniform (as this was the exogenously implemented ``prior of the prior"). Accordingly, the default beta distribution shown before eliciting the posterior was the participant's own prior report.} (Figure \ref{fig:Posterior example}). Once this was done, a new urn from the pool of 99 urns was selected, with replacement, and the process was repeated 30 times. All subjects faced the same set of urns and sequences of draws, but the order of tasks was randomized across participants. All belief-elicitation tasks were incentivized using a binarized scoring rule (more details are provided in Section \ref{sec.eliciting} and Appendix \ref{sec.scoring rules}).

To assess the role of preference-based biases, fifteen of the thirty tasks, placed at random positions in the experiment, required participants to report beliefs about urns to which an additional payment was attached. I henceforth refer to these as \textit{dollar urns}. In these tasks, participants received a payment in cents equal to the (unknown) number of red balls in the selected urn. Thus, participants had a monetary incentive to prefer urns with a higher proportion of red balls.

\subsection{Eliciting beta distributions}
\label{sec.eliciting}

To elicit belief distributions, I use the tool introduced in \cite{gonzalez2025direct}, which was validated there against non-parametric elicitation methods. This method is especially convenient in settings where beliefs are expected to follow a parametric form, while remaining closely connected to the broader distribution elicitation literature \citep{manski2004measuring,goldstein2008choosing,harrison2017scoring,crosetto2023comparing}.

\begin{figure}[!htbp]
     \centering
     \begin{subfigure}[b]{0.49\textwidth}
         \includegraphics[width=\textwidth]{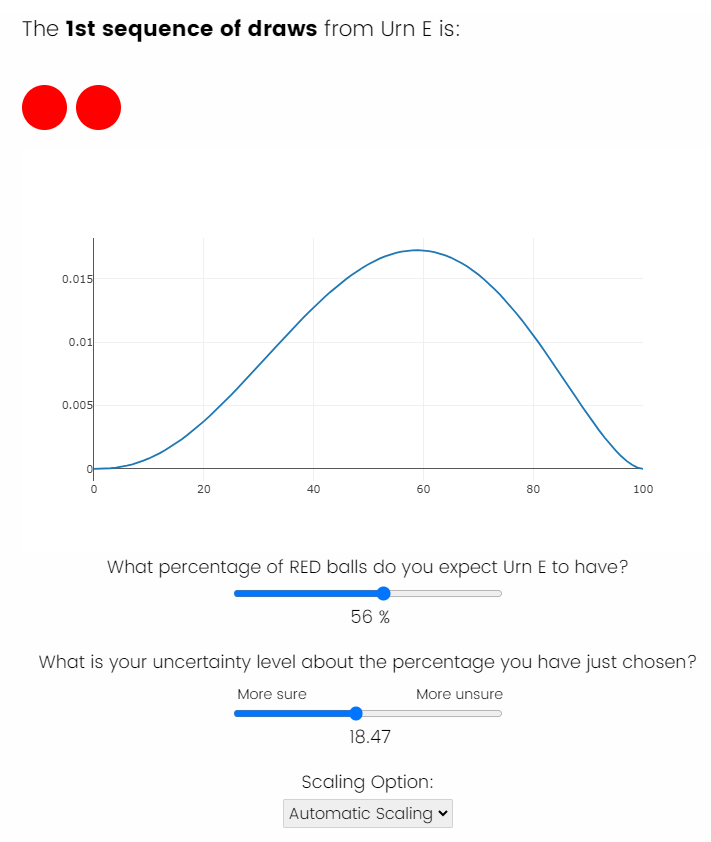}
         \caption{Figure 3a: Example of prior belief elicitation after observing a first sequence of draws from a random urn showing ``\textit{red, red}". This example shows a specific selection of a scaled beta distribution with an expected value of 56 and a standard deviation of 18.47.}
         \label{fig:Prior example}
     \end{subfigure}
     \hfill
     \begin{subfigure}[b]{0.49\textwidth}
         \includegraphics[width=\textwidth]{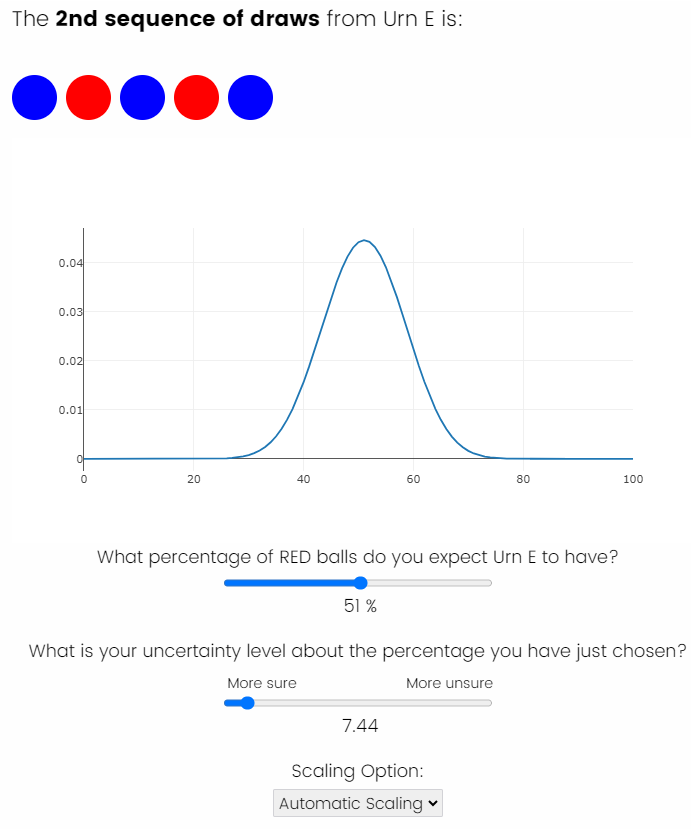}
         \caption{Figure 3b: Example of posterior belief elicitation after observing a second sequence of draws showing ``\textit{blue, red, blue, red, blue}". The updated beta distribution selected in this example has an expected value of 51 and a standard deviation of 7.44.}
         \label{fig:Posterior example}
     \end{subfigure}
\end{figure}

To report their beliefs, participants were presented with a dynamic graphical interface that allowed them to select their preferred beta distribution (see Figures \ref{fig:Prior example} and \ref{fig:Posterior example}). By moving two sliders, each associated with a different question, participants were able to select a specific beta distribution. These questions were:

{\itshape
\par
1. What percentage of red balls do you expect the selected urn to have?
\par
2. What is your uncertainty level about this percentage?
}

\vspace{5mm}

The first slider, associated with question (1), allows subjects to manipulate the expected value of a beta distribution, while the second slider, associated with question (2), allows them to manipulate its standard deviation.\footnote{The probability density function of any beta distribution can also be parameterized by its expected value and variance.} Participants were shown a five-minute explanatory video on how to interpret the graph they selected and how to manipulate it. Importantly, while participants adjusted sliders to report their beliefs, the relevant object for the analysis is the belief distribution they constructed, which was later mapped into beta-distribution parameters, rather than the individual slider positions themselves. Subjects were specifically instructed to solve the task graphically, and were also shown that the scale of the graph updates dynamically in order to keep the plot informative. Participants additionally had the option to fix the graph on a constant scale if they wished to do so. Before beginning the relevant belief-elicitation tasks, participants answered related comprehension questions\footnote{In order to participate in the experiment, subjects had to answer at least 3 out of 5 comprehension questions correctly. If any comprehension question was answered incorrectly, participants were given a second chance to answer all questions correctly.} and were allowed to test the software. For ease of interpretation, the interface did not allow subjects to report bimodal beta distributions. Bayesian reporting is nevertheless always feasible.\footnote{Standard ex-post parametric fitting of the beta distribution imposes the same restriction \citep{engelberg2009comparing}.}

More generally, because of the continuous nature of the variance slider, measurement error in uncertainty reports may affect bias identification. Appendix \ref{subsec.Measurement_Error} therefore applies a noise-injection procedure to assess the robustness of the results to measurement error in variance reports. This procedure imposes no restriction on the admissible variance range to rule out bimodality.

To incentivize truthful reporting, a binarized scoring rule was implemented in every task. Specifically, I follow a method similar to that suggested by \citet{schlag2013eliciting}, who propose scoring rules to incentivize different moments of a probability distribution (see Appendix \ref{sec.scoring rules} for the specific rule used here). At the same time, I follow \citet{danz2022belief} in not explicitly disclosing the exact scoring rule to participants. Instead, participants were told that, in order to maximize their expected payoff, they should always truthfully report both their guess for the percentage of red balls in the selected urn and their uncertainty about that percentage (that is, the mean and standard deviation of the beta distribution they selected).

\section{Baseline and Complete model regressions}
\label{sec.methodology}

This section presents the empirical specifications derived from the theoretical framework in Section \ref{sec.Theory}. Section \ref{sec.Meth.baseline} introduces a baseline regression that captures only a limited set of belief biases. Section \ref{sec.Meth complete} then presents the complete specification (equations \eqref{eq.final a}, \eqref{eq.final b} and \eqref{eq.final Var}), which incorporates the full set of biases studied in the model. Section \ref{sec.Meth Comp} introduces a procedure to compare the relative importance of the biases at the individual level. Additionally, I also estimate all intermediate model specifications to examine which bias classifications change when particular competing distortions are introduced. This fuller exercise makes it possible to assess which detected biases remain, and which biases lose significance once an increasingly richer bias set is taken into account. Models are estimated at the population and individual levels, since aggregate estimates may conceal substantial heterogeneity across subjects. Finally, Appendix \ref{sec.Grether_Check} reports a Grether-style reduced-form specification of the data in the traditional binary-state framework used in the belief-updating literature.

\subsection{Baseline model regressions}
\label{sec.Meth.baseline}

The baseline model closely resembles equations \eqref{eq.non-bayes S_a} and \eqref{eq.non-bayes S_b}. It tests for under(over)-inference and base-rate neglect(overuse): 
\begin{equation}
\label{eq.baseline a}
    \Tilde{a}_n-1 = \gamma_s k + \delta_s (a_0 - 1) + \varepsilon_a
\end{equation}
\begin{equation}
\label{eq.baseline b}
    \Tilde{b}_n-1 = \gamma_f (n -k) + \delta_f (b_0 - 1) + \varepsilon_b
\end{equation}
In equations \eqref{eq.baseline a} and \eqref{eq.baseline b}, successes and failures (variables $k$ and $(n-k)$), represent the number of red and blue balls observed in the second sequence of draws provided to participants. Variables $a_0$ and $b_0$ are the parameters of the beta distribution reported by subjects after the first sequence of draws, while $\Tilde{a}_n$ and $\Tilde{b}_n$ are the parameters of the beta distribution reported after observing the second sequence of draws. $\varepsilon_a$ and $\varepsilon_b$ are the error terms.

The parameter interpretation of equations \eqref{eq.baseline a} and \eqref{eq.baseline b} is akin to equations \eqref{eq.non-bayes Conf_alpha} and \eqref{eq.non-bayes Conf_beta }. Parameters $\gamma$ and $\delta$ similarly indicate under/over inference and base-rate neglect/overuse respectively. However, equations \eqref{eq.baseline a} and \eqref{eq.baseline b} differ by acknowledging that these biases ($\gamma_s$,$\gamma_f$) and ($\delta_s$,$\delta_f$) may vary between successes and failures (i.e. realizations of red and blue balls). The presence of those biases is tested by comparing whether the estimated $\gamma$ and $\delta$ parameters are significantly different from their Bayesian values (i.e. $\gamma_s=\gamma_f=1$ and $\delta_s=\delta_f=1$). 

Note that equations \eqref{eq.baseline a} and \eqref{eq.baseline b} do not have an intercept. This has a very straightforward theoretical justification. If all the independent variables in those equations take value $0$ (i.e. $k=0, n=0, a_0=1,b_0=1 $), both $\Tilde{a}_n$ and $\Tilde{b}_n$ take a value of $1$. This means that when one starts updating from a uniform prior distribution ($a_0=1,b_0=1 $), and observes no information signals whatsoever ($k=0,n=0$), one must not update, i.e. remain at such uniform prior after updating ($\Tilde{a}_n=1,\Tilde{b}_n=1$). Having an intercept different than zero would imply that agents update in the absence of any kind of information.

\subsection{Complete model regressions}
\label{sec.Meth complete}
In order to test for the presence of all the biases described in section \ref{sec.model multiple}, I run a slightly modified version of equations \eqref{eq.non-bayes Conf_alpha}, \eqref{eq.non-bayes Conf_beta } and \eqref{eq.Var}. Namely:
\begin{equation}
\label{eq.final a}
    \Tilde{a}_n - 1 = (\alpha_{0} + \alpha_{Pref}I_{Pref} + \alpha_{Seq}I_{Seq_s}) k + \rho_s  c + \delta_s (a_0 - 1) + \varepsilon_a
\end{equation}
\begin{equation}
\label{eq.final b}
    \Tilde{b}_n - 1 = (\beta_{0} + \beta_{Pref}I_{Pref} + \beta_{Seq}I_{Seq_f}) (n -k) + \rho_f  c + \delta_f (b_0 - 1) + \varepsilon_b
\end{equation}
\begin{equation}
\label{eq.final Var}
\Tilde{Var_n} = \eta + \nu \times Var_n + \varepsilon_v
\end{equation}
Variables in equations \eqref{eq.final a} and \eqref{eq.final b} are identical to the baseline model with the only exception of variable $c$, the relative measure of confirmation, as described in section \ref{sec. Confirmation}. Equation \eqref{eq.final Var} is almost identical to equation \eqref{eq.Var} in the interpretation of its variables and the $\nu$ parameter indicating over or underprecision ($\nu<1$ vs $\nu>1$). The inclusion of an intercept $\eta$ and an error term $\varepsilon_v$ are the only differences.\footnote{Similarly to the baseline model, equations \eqref{eq.final a} and \eqref{eq.final b} do not have an intercept for the very same reason as outlined in section \ref{sec.Meth.baseline}.}

Parameters $\rho$ and $\delta$ in equations \eqref{eq.final a} and \eqref{eq.final b} capture confirmation biases, and base-rate distortions respectively. As in the baseline case, equations \eqref{eq.final a} and \eqref{eq.final b} differ by acknowledging that these parameters may differ between successes and failures ($\rho_s$ vs $\rho_f$) and ($\delta_s$ vs $\delta_f$). 

Another important difference between equations \eqref{eq.final a}, \eqref{eq.final b} and \eqref{eq.non-bayes Conf_alpha},\eqref{eq.non-bayes Conf_beta } lies in accommodating the biases of section \ref{sec.biases seq and pref}. This involves the inclusion of three dummy variables – $I_{Pref}$, $I_{Seq_s}$, and $I_{Seq_f}$ – interacting with the number of successes (in \eqref{eq.final a}) or failures (in \eqref{eq.final b}). $I_{Pref}$ equals 1 when the subject faces a decision with preferences over the state-space (that is, when beliefs about a \textit{dollar urn} were reported) and 0 otherwise. $I_{Seq_s}$ ($I_{Seq_f}$) equals 1 when the last three balls observed in the second sequence are red (blue). By doing so, one can distinguish preference-based biases (optimism and pessimism), sequence-related biases (gambler's fallacy and hot-hand fallacy) and over/under inference. The presence of those biases is tested by comparing whether the estimated parameters are significantly different from their Bayesian values. Importantly, this means that the interaction coefficients are tested against zero, while the baseline information weights are tested against one. For example, identifying a sequence-related bias requires evidence that a streak changes information weighting beyond the baseline response (over/under inference) to successes or failures.  Table \ref{tab:summary biases} summarizes the baseline and complete models, and specifically shows the ranges of values of the different parameters, which correspond to each bias.

\begin{table}[ht]
\centering\scriptsize
\begin{tabular}{p{6cm} p{9cm}}
\toprule
\textbf{Baseline Model (Eqs. 12, 13)} & \textbf{Complete Model (Eqs. 14, 15, 16)} \\
\midrule
\multicolumn{2}{l}{\textbf{Inference Biases}} \\
$\gamma_s$ or $\gamma_f > 1$: Overinference & $\alpha_0$ or $\beta_0 > 1$: Overinference \\
$\gamma_s$ or $\gamma_f \in [0,1)$: Underinference & $\alpha_0$ or $\beta_0 \in [0,1)$: Underinference \\
$\gamma_s$ or $\gamma_f <0$: Against signal &
$\alpha_0$ or $\beta_0 <0 $: Against signal \\
\midrule
\multicolumn{2}{l}{\textbf{Base-Rate Biases}} \\
$\delta_s$ or $\delta_f > 1$: Base-Rate Overuse & $\delta_s$ or $\delta_f > 1$: Base-Rate Overuse \\
$\delta_s$ or $\delta_f < 1$: Base-Rate Neglect & $\delta_s$ or $\delta_f < 1$: Base-Rate Neglect \\
\midrule
\multicolumn{2}{l}{\textbf{Confirmation Biases}} \\
 & $\rho_s$ or $\rho_f < 0$: Confirmation Bias \\
 & $\rho_s$ or $\rho_f > 0$: Disconfirmation Bias \\
\midrule
\multicolumn{2}{l}{\textbf{Preference-Based Biases}} \\
 & $\alpha_0 + \alpha_{Pref} > 1$ or $\beta_0 + \beta_{Pref} < 1$: Optimism \\
 & $\alpha_0 + \alpha_{Pref} < 1$ or $\beta_0 + \beta_{Pref} > 1$: Pessimism \\[3pt]
 & \textbf{Asymmetries:}\\[3pt]
 & $\alpha_{Pref} > \beta_{Pref}$: Good News Effect \\
 & $\alpha_{Pref} < \beta_{Pref}$: Bad News Effect \\
\midrule
\multicolumn{2}{l}{\textbf{Sequence-related Biases}} \\
 & $\alpha_0 + \alpha_{Seq} > 1$ or $\beta_0 + \beta_{Seq} > 1$: Hot-Hand Fallacy \\
 & $\alpha_0 + \alpha_{Seq} < 1$ or $\beta_0 + \beta_{Seq} < 1$: Gambler's Fallacy \\
\midrule
\multicolumn{2}{l}{\textbf{Precision Biases}} \\
 & $\nu > 1$: Underprecision \\
 & $\nu < 1$: Overprecision \\
\bottomrule
\end{tabular}
\caption{Summary of biases for Baseline and Complete Model}
\label{tab:summary biases}
\par\smallskip
\begin{minipage}{\linewidth}
\footnotesize
\textit{Note:} Additionally, for bias classification, sequence and preference interaction coefficients ($\alpha_{Pref},\beta_{Pref},\alpha_{Seq}, \beta_{Seq}$) are also tested against zero; baseline coefficients are tested against one.
\end{minipage}
\end{table}


\paragraph{Intermediate model regressions} The baseline and complete specifications also allow for a systematic comparison of intermediate models. Starting from the baseline equations, confirmation effects, preference-based biases, and sequence-related biases can each be introduced separately or in combination. These three additional families yield $2^3=8$ specifications (see Table \ref{tab:intermediate_models}), including the baseline and complete models. An omitted family is excluded by setting its corresponding coefficients in equations \eqref{eq.final a} and \eqref{eq.final b} to zero, while all remaining coefficients are re-estimated. Comparing these specifications makes it possible to examine how the identification of one bias depends on whether other distortions are allowed for.\footnote{Precision biases are estimated through the separate variance equation \eqref{eq.final Var}, whose specification is unchanged across these comparisons.}

\begin{table}[!htbp]
\centering \scriptsize
\begin{tabular}{lccc}
\toprule
\textbf{Model} & \textbf{Confirmation} & \textbf{Preference-based}
              & \textbf{Sequence effects} \\
\midrule
Baseline       & No  & No  & No  \\
Intermediate 1 & Yes & No  & No  \\
Intermediate 2 & No  & Yes & No  \\
Intermediate 3 & Yes & Yes & No  \\
Intermediate 4 & No  & No  & Yes \\
Intermediate 5 & Yes & No  & Yes \\
Intermediate 6 & No  & Yes & Yes \\
Complete       & Yes & Yes & Yes \\
\bottomrule
\end{tabular}
\caption{Bias families included in each specification.}
\label{tab:intermediate_models}
\par\smallskip
\begin{minipage}{0.95\linewidth}
\footnotesize
\textit{Note:} All specifications include inference and base-rate
biases. Precision biases are estimated through the separate
variance equation \eqref{eq.final Var}.
\end{minipage}
\end{table}

\subsection{The effect of each individual bias}
\label{sec.Meth Comp}

A bias may be frequently detected across subjects and yet have only a modest quantitative effect on inference when it appears. Conversely, a less prevalent bias may generate large distortions in posterior beliefs. For this reason, beyond classifying which biases are present, I also construct a measure of the relative importance of each bias in driving deviations from the Bayesian benchmark. This measure identifies the specific effect of each bias in the changes of mean and variance with respect to their Bayesian counterpart. In order to achieve this, I compare the Bayesian expected value and Bayesian variance to bias-specific counterfactual moments that isolate the effect of one particular bias. I refer to these objects as \textit{Bias-specific expected value} and \textit{Bias-specific variance}. The Bayesian expected value and variance of the posterior beta distribution are given by: 
\begin{align}
E_n &\equiv E(p|a_n,b_n) = \frac{a_n}{a_n + b_n} \\
Var_n &\equiv Var(p|a_n,b_n)=\frac{a_n b_n}{(a_n + b_n)^{2}(a_n + b_n +1)}
\end{align}
The bias-specific expected value $E_{Bias}$ and bias-specific variance $Var_{Bias}$\footnote{In the case of precision biases, $Var_{Bias}$ is already given by $\Tilde{Var}_{n}$, and $E_{Bias}$ would be unaffected.} are given by:
\begin{align}
E_{Bias} &= \frac{a_{bias}}{a_{bias} + b_{bias}} \\
Var_{Bias} &= \frac{a_{bias} b_{bias}}{(a_{bias} + b_{bias})^{2}(a_{bias} + b_{bias} +1)}
\end{align}
where $a_{bias}$ and $b_{bias}$ are the hypothetical parameters of a posterior beta distribution if they were distorted by only one specific bias.\footnote{Because we are looking at the effect of specific biases for each subject, it is theoretically possible that the effect of the bias in isolation is so strong that it makes $a_{bias}$ or $b_{bias}$ negative. This would make $E_{Bias}$ and $Var_{Bias}$ uninterpretable. Therefore, it is assumed that the maximum effect that a bias in isolation can have is such that the associated parameter $a_{bias}$ or $b_{bias}$ is equal to 0 + $\varepsilon$. It is worth noting that these cases were very rare, as it would require strongly updating against the information signal.} For example, suppose we want to evaluate the impact of \textit{overinference}. Suppose a particular subject exhibits \textit{overinference} only in successes: That is $\alpha_0>1$. Then, $a_{bias}=a_{Overinference}=\alpha_0k +a_0$ and $b_{bias}=b_n$. Equations (19) and (20) for every other bias are calculated analogously. 

To capture the relative importance of each bias in inference, I compare the distance between the Bayesian and distorted expected value and variance. Namely:
\begin{align}
    \Delta E_{Bias} &= |E_n - E_{Bias}| \\
    \Delta Var_{Bias} &= |Var_n - Var_{Bias}|
\end{align}

Equations (21) and (22) allow us to quantify the discrepancy between the Bayesian expected value and variance, and the bias-specific measures. These discrepancies serve as indicative measures of the influence of each bias on the overall inference process.

\section{Results}
\label{sec: Results}
\subsection{Population-level analysis}

Table \ref{tab:pop} reports the baseline model (equations \eqref{eq.baseline a} and \eqref{eq.baseline b}) and the complete model (equations \eqref{eq.final a}, \eqref{eq.final b} and \eqref{eq.final Var}) estimated at the population level. Columns (1) and (2) correspond to the baseline model, while columns (3), (4) and (5) correspond to the complete model. All regressions in Table \ref{tab:pop} are estimated with standard errors clustered at the participant level.\footnote{A negligible fraction of observations (5 out of 2640, or 0.2\%) yielded slightly negative values for $a_0$ and $b_0$ due to a minor margin of error in the graphical interface. This only occurred in rare cases where the reported variance was exceptionally high and the expected value was near the extremes (0.01 or 0.99). These observations were excluded from the regressions.}


\begin{table}[!htbp]\centering\scriptsize
\begin{tabular}{lccccc}\hline
 & $a_n-1$ & $b_n-1$ & $a_n-1$ & $b_n-1$ & Posterior variance \\
 & (1) & (2) & (3) & (4) & (5) \\ \hline
Successes & 37.234$^{***}$ &  & 1.773 &  &  \\
 & (11.673) &  & (42.084) &  &  \\
Prior a $-1$  & 0.017$^{***}$ &  & 0.005$^{***}$ &  &  \\
 & (0.020) &  & (0.026) &  &  \\
Failures &  & 76.207$^{**}$ &  & 69.388 &  \\
 &  & (32.010) &  & (60.270) &  \\
Prior b $-1$  &  & -1.78e-04$^{***}$ &  & -5.50e-04$^{***}$ &  \\
 &  & (0.019) &  & (0.019) &  \\
Successes: preference &  &  & -30.655 &  &  \\
 &  &  & (27.861) &  &  \\
Successes: positive sequence &  &  & 14.219 &  &  \\
 &  &  & (32.960) &  &  \\
Failures: preference &  &  &  & -26.843 &  \\
 &  &  &  & (71.736) &  \\
Failures: negative sequence &  &  &  & 104.483 &  \\
 &  &  &  & (101.168) &  \\
Confirmation measure &  &  & 468.800 & -31.905 &  \\
 &  &  & (300.870) & (75.693) &  \\
Bayesian variance &  &  &  &  & 0.992 \\
 &  &  &  &  & (0.077) \\
Constant &  &  &  &  & 0.002$^{**}$ \\
 &  &  &  &  & (0.001) \\
\hline
Observations & 2635 & 2635 & 2635 & 2635 & 2635 \\
$R^2$ & 0.002 & 0.003 & 0.003 & 0.004 & 0.292 \\
Adjusted $R^2$ & 0.001 & 0.002 & 0.002 & 0.002 & 0.292 \\
\hline\end{tabular}
\caption{Baseline and complete models at the population level. Standard errors are clustered by participant. Stars denote tests against the Bayesian benchmark: $^{*}p<0.1$, $^{**}p<0.05$, $^{***}p<0.01$.}
\label{tab:pop}
\end{table}


As in the rest of the paper, significance is defined relative to the corresponding Bayesian benchmark rather than relative to zero (see Table \ref{tab:summary biases}). In the baseline specification, two biases are significant at the population level: overinference and base-rate neglect. In particular, the coefficients on successes and failures are significantly above their Bayesian benchmarks, while the coefficients on prior beliefs are significantly below one. Thus, when the data are summarized at the aggregate level, the baseline model suggests that subjects overweight the information signal and underweight prior beliefs.

Once the full set of biases is incorporated, the evidence for population-level overinference disappears, whereas base-rate neglect remains significant. This is the main comparison between the baseline and complete specifications at the aggregate level: when a richer set of biases is factored in, baseline overinference estimates appear to be partly absorbing other distortions. In addition, Appendix \ref{sec.Grether_Check} reports a Grether-style reduced-form specification of the same data in the traditional binary-state framework. The resulting estimates are qualitatively consistent both with the classic belief-updating literature and with the broad aggregate patterns reported here.

At the same time, Table \ref{tab:pop} shows that the population-level equations explain little of the variation in posterior reports. The goodness of fit is extremely low in columns (1) to (4), for both the baseline and complete specifications. However, this need not be interpreted as the model being unable to explain subjects' updating behavior. Instead, it may reflect the fact that different subjects exhibit different distortions, which tend to average out in the population regression. Consistent with this interpretation, Table \ref{tab:R2 comp} in Appendix \ref{sec.extra} shows that the goodness of fit of all equations improves markedly when the model is estimated separately for each participant, with the complete model at the individual level showing the best fit. At the same time, Table \ref{tab:info crit} shows analogous improvements in Akaike and Bayesian Information Criteria. 

These results point to substantial individual heterogeneity in belief-updating distortions and motivate the individual-level analysis that follows.

\subsection{Individual-level analysis and bias misattribution}
\label{sec.ind analysis}

\begin{figure}[ht]
     \centering
     \begin{subfigure}[a]{0.49\textwidth}
         \includegraphics[width=\textwidth]{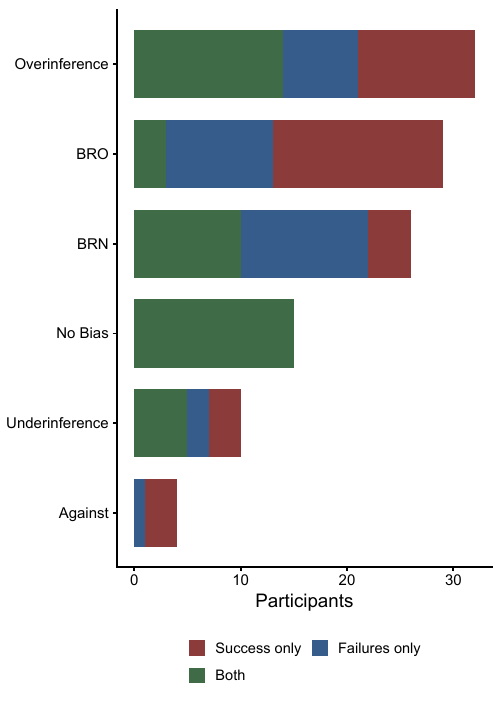}
         \caption{Figure 4a: Number of times a specific bias is found to be significant ($p<0.05$) in the baseline model at the individual level.}
         \label{fig:count baseline}
     \end{subfigure}
     \hfill
     \begin{subfigure}[a]{0.49\textwidth}
         \includegraphics[width=\textwidth]{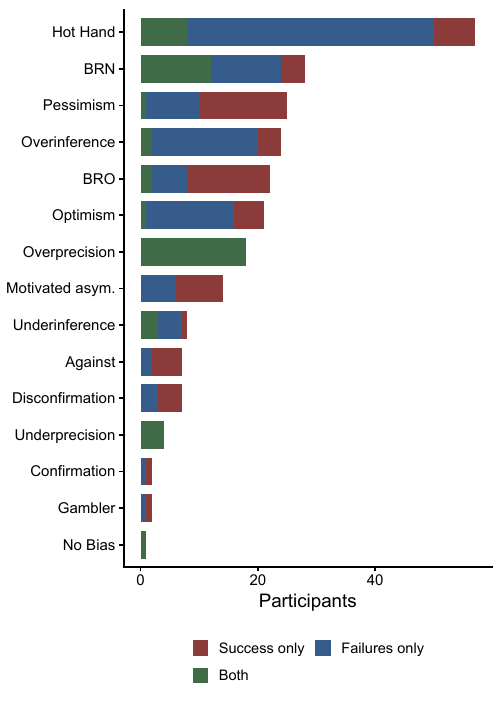}
         \caption{Figure 4b:  Number of times a specific bias is found to be significant ($p<0.05$) in the complete model at the individual level.}
         \label{fig:count complete}
     \end{subfigure}
\end{figure}

Figures \ref{fig:count baseline} and \ref{fig:count complete} provide an overview of the biases detected at the individual level within the context of the two models: the baseline and the complete model. In Figure \ref{fig:count baseline}, the bar chart depicts the occurrences of statistically significant biases at the 5\% level within the baseline model. The color-coded representation distinguishes biases: red means significance solely for successes, blue for failures, and green for significance in both successes and failures. Notably, overinference and base-rate overuse emerge prominently. Figure \ref{fig:count complete}, illustrating individual biases in the complete model, shows a more complete picture. 

Here, a broader array of biases is accounted for, with the hot-hand fallacy being the most commonly exhibited bias among the subjects.\footnote{This is especially surprising given that sequence-related biases, such as the hot-hand fallacy, are tested with a reduced number of observations, which generally limits statistical power and thus makes significant findings less expected.} Further comparing these models yields interesting insights. First, at the descriptive level, it becomes evident that individual heterogeneity persists across both models: Each tested bias category is detected for at least some subjects, although prevalence differs substantially across biases. Notably, the prevalence of the hot-hand fallacy contrasts strongly with the sparsity of confirmatory biases, underprecision or gambler's fallacy among subjects. Second, moving from the baseline to the complete model substantially reduces the set of subjects who remain uncategorized.

While the ``No Bias'' bar in Figure \ref{fig:count baseline} shows that 15 individuals (17\% of the sample) could not be categorized --that is, they are either too noisy to classify or statistically indistinguishable from Bayesian updaters-- in the complete model this number falls to one. The prevalence of overinference and base-rate overuse also diminishes when transitioning from the baseline to the complete model.

These changes raise a central question: do departures from Bayesian updating continue to be attributed to the same bias once other biases have been incorporated into the model? Since the complete model introduces several bias families simultaneously, the baseline-to-complete comparison is not enough to distinguish their respective roles. The intermediate specifications in Table \ref{tab:intermediate_models} make it possible to examine this question by introducing one family at a time. More generally, they reveal which bias classifications remain present and which lose significance once competing distortions are progressively taken into account.

\subsubsection*{Bias Misattribution}

To address this question, I compare the eight specifications by introducing one additional bias family at a time. This yields twelve comparisons between nested models (i.e. baseline to intermediate, intermediate to intermediate, and intermediate to complete). Each comparison uses the same number of observations and retains the significance tests and classification rules described in Table~\ref{tab:summary biases}. Let $B_{ij}(m)$ equal one if subject $i$ is classified as exhibiting bias $j$ under specification $m$, and zero otherwise. Let $\mathcal{C}_j$ denote the set of comparisons $(m,m')$ in which $m'$ adds one family to $m$ and bias $j$ can be identified in both specifications. I define the classification-loss rate as
\begin{equation}
\widehat{L}_j =
\frac{
\displaystyle\sum_{(m,m')\in\mathcal{C}_j}
\sum_{i=1}^{N}
B_{ij}(m)\bigl[1-B_{ij}(m')\bigr]
}{
\displaystyle\sum_{(m,m')\in\mathcal{C}_j}
\sum_{i=1}^{N}
B_{ij}(m)
}
\label{eq:classification_loss}
\end{equation}
This measure gives the fraction of initially detected classifications that disappear after another family is introduced. A higher percentage shows that the identification of a bias is more sensitive to the inclusion of competing distortions.\footnote{A classification may disappear because the relevant coefficients lose significance or because the estimated parameters no longer satisfy the conditions for that particular bias. The measure therefore records sensitivity of classification; it does not establish that every classification lost under an expanded model was initially incorrect. The main analysis uses the same significance threshold, $p<0.05$, throughout.}

\begin{figure}[ht]
    \centering
    \captionsetup[subfigure]{
        font=scriptsize,
        justification=centering,
        skip=4pt
    }

    \begin{subfigure}[t]{0.43\textwidth}
        \centering
        \includegraphics[width=\linewidth]
            {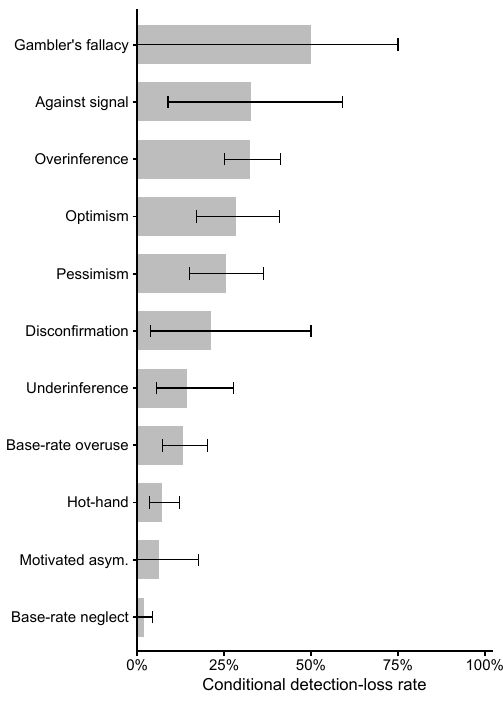}
        \caption{Classification-loss rates across competing specifications.}
        \label{fig:misattribution_sensitivity}
    \end{subfigure}
    \hfill
    \begin{subfigure}[t]{0.56\textwidth}
        \centering
        \includegraphics[width=\linewidth]
            {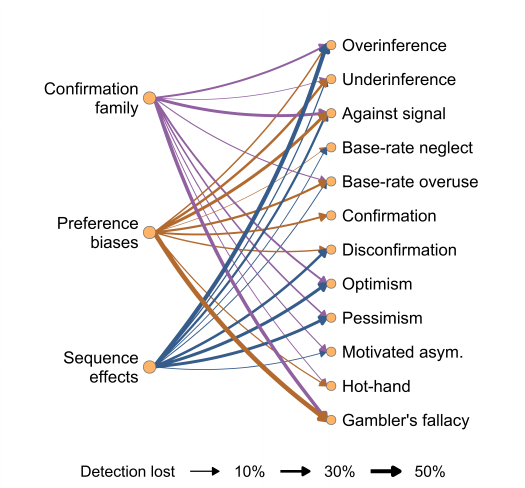}
        \caption{Classification losses by added bias family.}
        \label{fig:misattribution_network}
    \end{subfigure}

    \caption{Sensitivity of bias classifications to the inclusion
    of competing distortions.}
    \label{fig:misattribution}

    \par\smallskip
    \begin{minipage}{0.98\textwidth}
        \footnotesize
        \textit{Notes:} Panel (a) reports the fraction of initially
        detected subject--model classifications that disappear when one additional bias family is introduced. Comparisons include
        only specifications containing the target bias both before
        and after the addition. Error bars show 95\% percentile bootstrap confidence intervals
        from 1,000 resamples of subjects, retaining each subject's
        estimated classifications across all models together. Biases exhibited by three or fewer subjects are left out of the figure. In panel (b), arrows run from the added family to the bias
        whose classification is lost. Arrow thickness represents the
        corresponding family-specific loss rate; colours distinguish
        the added families.
    \end{minipage}
\end{figure}

Figure \ref{fig:misattribution_sensitivity} shows substantial differences across biases. Overinference loses its classification in 32.5\% of the comparisons in which it is initially detected. By contrast, the corresponding rates are 7.2\% for hot-hand fallacy and only 1.9\% for base-rate neglect. The prominence of these latter biases therefore survives the inclusion of other sources of distorted updating. In particular, hot-hand fallacy remains both widespread and comparatively stable across the specifications in which it can be identified.\footnote{Gambler's fallacy has the highest classification-loss rate, at 50\%, but this corresponds to six losses among twelve eligible comparisons involving only five subjects. Results for such infrequent biases are consequently much less precisely determined.}

Figure \ref{fig:misattribution_network} shows which additional families produce these changes. An arrow points from the family introduced into the model to the bias whose classification is lost, with its thickness indicating the corresponding conditional loss rate. This figure allows us to see the source of this loss. Sequence effects account for the largest loss rate in overinference. I find that across the four comparisons that introduce this family, 75 of 143 initially detected overinference classifications disappear (52.4\%). Similar to the previous discussion from simply comparing the baseline and the complete model, a substantial part of what appears to be general overinference in a restricted specification is sensitive to allowing subjects' responses to depend on the sequence of observed signals.\footnote{Importantly, the coefficient diagnostics reported in Figure \ref{fig:misattribution_coefficient_se} in Appendix \ref{sec.extra} strongly support this interpretation. Among inference tests that lose significance, movement of the estimated coefficients toward the Bayesian benchmark accounts for a substantially larger average contribution to the decline in statistical significance than changes in standard errors.}

Another strong interesting pattern is that preference-based classifications are also sensitive to the inclusion of other biases. Optimism and pessimism have classification-loss rates of 28.6\% and 25.7\%, respectively, where sequence effects once again produce the largest losses. In just over half of the losses for each category, significant preference effects remain after the original classification disappears. 

The contrast between overinference and the more stable hot-hand and base-rate-neglect classifications persists under multiple-hypothesis correction. Applying the same adjusted threshold, $p<0.05/12$, to all eight specifications yields classification-loss rates of 53.7\% for overinference, 2.0\% for hot-hand fallacy, and 2.8\% for base-rate neglect. When sequence effects are introduced, the loss rate for overinference reaches 70.3\%. These comparisons reinforce the empirical importance of allowing for competing distortions: conclusions about overinference depend substantially on the specification, while hot-hand fallacy and base-rate neglect remain comparatively stable.

Appendix \ref{sec.robust} revisits these comparisons under a model-specific multiple-hypothesis testing correction. The main conclusions remain largely unchanged: There is substantial individual heterogeneity across all bias categories, the hot-hand fallacy is the most commonly exhibited bias, and overinference, optimism and pessimism are very sensitive to the inclusion of sequence effects. Interestingly, the confirmation-bias family disappears in the complete model after imposing a much more demanding significance threshold.

\subsection{Quantitative importance of biases: First and second moment effects}

\begin{figure}[!ht]
    \centering
    \captionsetup[subfigure]{
        font=scriptsize,
        justification=centering,
        singlelinecheck=false,
        skip=3pt
    }

    \begin{subfigure}[t]{0.48\textwidth}
        \centering
        \includegraphics[
            width=\linewidth,
            height=0.28\textheight,
            keepaspectratio
        ]{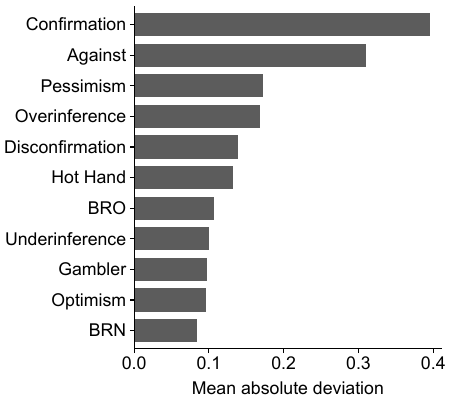}
        \caption{Expected value deviations from Bayesian framework: conditional average.}
        \label{fig:inference EV gross}
    \end{subfigure}
    \hfill
    \begin{subfigure}[t]{0.48\textwidth}
        \centering
        \includegraphics[
            width=\linewidth,
            height=0.28\textheight,
            keepaspectratio
        ]{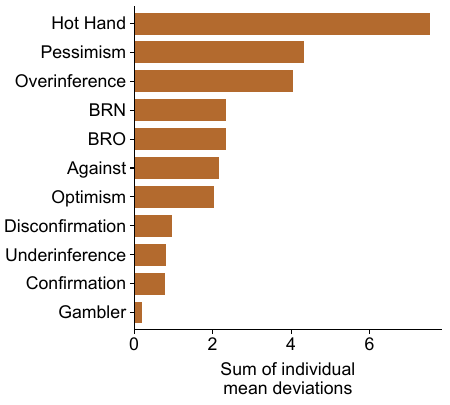}
        \caption{Expected value deviations from Bayesian framework: adjusted for prevalence.}
        \label{fig:inf EV net}
    \end{subfigure}

    \par\vspace{0.6em}

    \begin{subfigure}[t]{0.48\textwidth}
        \centering
        \includegraphics[
            width=\linewidth,
            height=0.28\textheight,
            keepaspectratio
        ]{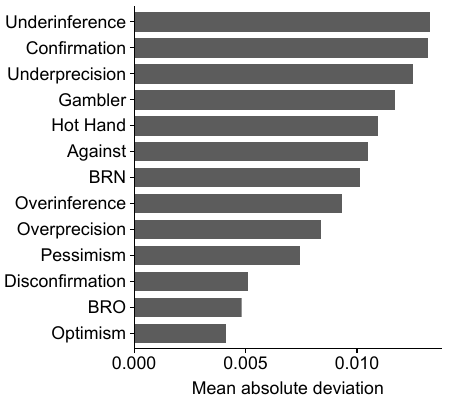}
        \caption{Variance deviations from Bayesian framework: conditional average.}
        \label{fig:inf VAR gross}
    \end{subfigure}
    \hfill
    \begin{subfigure}[t]{0.48\textwidth}
        \centering
        \includegraphics[
            width=\linewidth,
            height=0.28\textheight,
            keepaspectratio
        ]{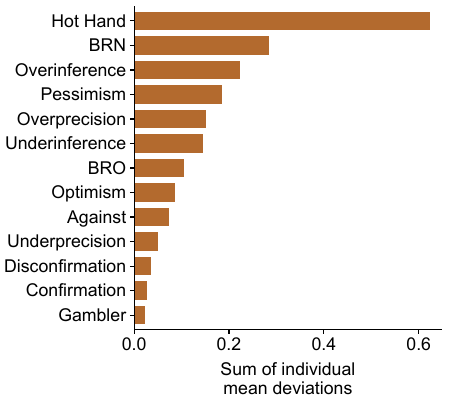}
        \caption{Variance deviations from Bayesian framework: adjusted for prevalence.}
        \label{fig: inf VAR net}
    \end{subfigure}

    \caption{Bias-specific deviations from Bayesian posterior
    expected values and variances.}
    \label{fig:bias_impacts}

    \par\smallskip
    \begin{minipage}{0.97\textwidth}
        \footnotesize
        \textit{Notes:} Left panels report average absolute deviations
        conditional on exhibiting each bias. Right panels sum
        individual mean absolute deviations across subjects exhibiting
        that bias.
    \end{minipage}
\end{figure}

Although these comparisons establish the prevalence of different biases, they do not tell us which biases matter most for inference. A bias may be common but quantitatively modest, or relatively less frequent but highly consequential when it appears. Thus, I apply the methodology outlined in Section \ref{sec.Meth Comp} to assess the relative importance of each bias. Figure \ref{fig:bias_impacts} reports deviations from Bayesian posterior expected values and variances within the complete model: The left panels present average deviations conditional on exhibiting each bias, while the right panels adjust for prevalence by summing these deviations across subjects exhibiting that bias.

In Figure \ref{fig:inference EV gross}, two biases, confirmation bias and updating against the information signal, stand out with prominent expected value deviations ($\approx 0.4$ and $\approx 0.35$ distance units to the Bayesian posterior expected value respectively).\footnote{One could, nonetheless, discount the prominence of confirmation bias since it is very rarely exhibited. Also, naturally, updating against the information signal is a strong mistake in terms of direction, so a big effect in the Bayesian expected value is unsurprising.} In contrast, most other biases exhibit deviations that range from $0.1$ to $0.2$ distance units to the Bayesian posterior expected value. A particularly interesting observation is that base-rate neglect, despite being quite commonly observed, has a very modest effect on shifting the Bayesian expected value if exhibited.  After correcting for prevalence in Figure \ref{fig:inf EV net}, hot-hand fallacy, pessimism, and overinference stand out over the rest.\footnote{Figures \ref{fig:inf EV net} and \ref{fig: inf VAR net} should be interpreted ordinally.}

Figures \ref{fig:inf VAR gross} and \ref{fig: inf VAR net} are analogous to \ref{fig:inference EV gross} and \ref{fig:inf EV net}, but represent variance deviations with respect to the Bayesian framework instead. While the relative contribution of each bias appears to be quite homogeneous (Figure \ref{fig:inf VAR gross}), once we adjust for bias prevalence (Figure \ref{fig: inf VAR net}), the hot-hand fallacy shows its prominence, while base-rate neglect, overinference and pessimism come next. This reinforces the importance of pessimistic beliefs in overall inference departures, both in expected-value and variance deviations.

\subsection{Co-occurrence patterns across belief biases}
\label{sec:cooccurrence}

Biases rarely appear in isolation. To summarize heterogeneity in a compact and easily interpretable way, I study how the biases estimated in the complete model co-occur across subjects.\footnote{The analysis of how biases co-occur across subjects was not pre-registered in this exact form.} Let $j \in \{1,\dots,J\}$ index the set of bias labels considered in the complete model (e.g., over-/under-inference, base-rate distortions, motivated asymmetries, sequence-related biases, (over/under)precision, and confirmation/disconfirmation).
For each subject $i\in\{1,\dots,N\}$, I define a binary indicator $B_{ij} \in \{0,1\},$
which equals one if subject $i$ is classified as exhibiting bias $j$ (according to the individual-level estimation), and zero otherwise. Stacking these indicators yields an $N\times J$ subject-by-bias matrix $B$.

For each bias $j$, I compute its prevalence
\begin{equation}
    \hat \upsilon_j \;=\; \frac{1}{N}\sum_{i=1}^N B_{ij}
\end{equation}

\noindent,which is simply a subject-normalized score for Figure \ref{fig:count complete} (e.g. hot-hand: $\hat \upsilon \approx 0.65$, overinference $ \hat \upsilon\approx 0.27$, base-rate neglect $\hat \upsilon \approx 0.31$ etc.).\footnote{Because some biases are low-prevalence, associations involving these biases should be interpreted cautiously.}

To quantify co-occurrence between two binary bias indicators $B_{\cdot j}$ and $B_{\cdot j'}$, I compute the phi correlation coefficient, $\phi_{jj'} = \mathrm{corr}(B_{\cdot j}, B_{\cdot j'}),$ which coincides with the Pearson correlation computed on $\{0,1\}$-valued variables. Positive values indicate that the two biases tend to be jointly exhibited, while negative values indicate that they tend to be mutually exclusive.

\begin{figure}[ht]
\centering
\includegraphics[width=0.85\textwidth]{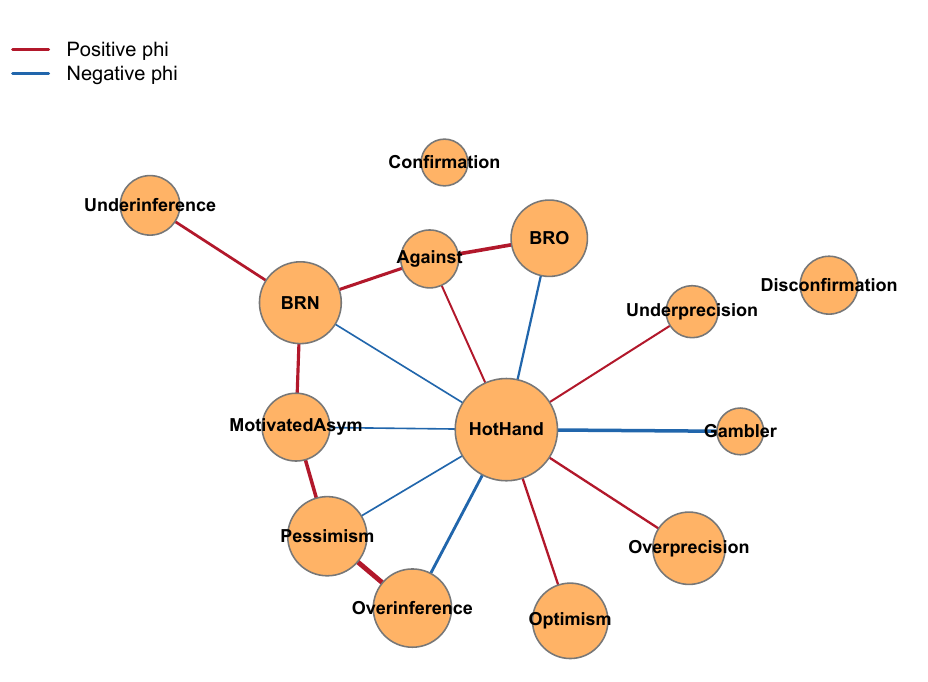}
\caption{Bias co-occurrence network. Nodes represent biases and node size is proportional to prevalence $\hat \upsilon_j$. A red edge is drawn \ when $\max\{\Pr(B_{ij}=1\mid B_{ij'}=1),\,\Pr(B_{ij'}=1\mid B_{ij}=1)\}\ge 0.5$. A blue edge is drawn when $\max\{\Pr(B_{ij}=1\mid B_{ij'}=0),\,\Pr(B_{ij'}=1\mid B_{ij}=0)\}\ge 0.5$, or  $\Pr(B_{ij'}=1\mid B_{ij}=1)\leq 0.2$ and $\Pr(B_{ij'}=1\mid B_{ij}=0)\geq 0.5$, for at least one ordering of $j$ and $j'$. Edge thickness is proportional to $|\phi_{jj'}|$.}
\label{fig:cooccur_network}
\end{figure}

Figure~\ref{fig:cooccur_network} provides a summary of the strongest co-occurrence patterns as a network between biases, while Figure \ref{fig:cooccur_heatmap} (in Appendix \ref{sec.extra}) provides a visual representation of the full correlation matrix. Nodes represent biases and node size is proportional to prevalence $\hat \upsilon_j$. An edge between two biases $j$ and $j'$ is drawn when their association is strong enough to have a transparent probabilistic interpretation. Specifically, I draw a red edge (positive co-occurrence) when observing one bias makes the presence of another bias more likely than not, i.e.\ when $\max\{\Pr(B_{ij}=1\mid B_{ij'}=1),\,\Pr(B_{ij'}=1\mid B_{ij}=1)\}\ge 0.5$. Conversely, I draw a blue edge (negative co-occurrence) when the reverse holds in the sense of near-mutual exclusivity, i.e.\ when the presence of bias $j$ makes it more unlikely that bias $j'$ is happening ( $\max\{\Pr(B_{ij}=1\mid B_{ij'}=0),\,\Pr(B_{ij'}=1\mid B_{ij}=0)\}\ge 0.5$). Edge thickness is proportional to $|\phi_{jj'}|$.\footnote{The co-occurrence criterion also requires that at least three participants exhibit both biases. The second criterion requires $\Pr(B_{ij'}=1\mid B_{ij}=1)\leq 0.2$ and $\Pr(B_{ij'}=1\mid B_{ij}=0)\geq 0.5$, for at least one ordering of $j$ and $j'$. Once edges are selected, their color follows the sign of $\phi_{jj'}$: red for positive and blue for negative correlations. A blue edge can also qualify through the co-occurrence criterion.}

Three patterns stand out. First, hot-hand fallacy is the most connected bias in the network, with ten links compared with four for base-rate neglect, the next most connected node. It also has the highest betweenness centrality, meaning that many of the shortest paths between other biases pass through it (see Table \ref{tab:network_centrality} in Appendix \ref{sec.extra}). Its connections extend across inference, prior weighting, preference-related biases and precision. These associations have different signs: overinference is detected in 23\% of participants who exhibit hot-hand fallacy, compared with 35\% of those who do not, while overprecision is detected in 23\% and 16\%, respectively. Hot-hand therefore appears across a broad range of updating patterns. This means that a participant exhibiting the hot-hand is somewhat predictive about which other distortions they also exhibits.

Second, removing hot-hand leaves a path-like component of red edges which links base-rate overuse, updating against the signal, base-rate neglect, motivated asymmetry, pessimism and overinference. Interestingly, underinference connects only with base-rate neglect: a co-exhibiting pattern which is consistent with the classic belief-updating literature. As Table \ref{tab:network_centrality} shows, base-rate neglect also has a second important connecting role within the network, with a high betweenness centrality. The strongest association along this chain is between pessimism and overinference ($\phi=0.35$): overinference is detected in 52\% of participants classified as pessimistic, compared with 17\% of the remaining participants. Thus, preference-related distortions frequently correlate with excessive weighting of information even outside the preference context. 

Finally, confirmation and disconfirmation have no connections meeting the edge-selection criteria. Together with their low prevalence in the individual-level analysis, this emphasizes the reduced importance of the confirmation family across most of the empirical findings. Because this analysis is partly exploratory, I treat these patterns as descriptive regularities rather than as definitive typologies.

\section{Concluding remarks}
\label{sec. conclusion}

This paper develops a unified framework for separately identifying multiple belief-updating biases that may otherwise be empirically confounded. The key methodological idea is to study belief updating using prior and posterior belief distributions over a continuous state space, rather than point beliefs alone. This allows a broader range of conflicting distortions to be brought into the same structural framework and tested in a laboratory setting using a novel belief-elicitation method.

The results show that accounting for a richer set of biases changes the interpretation of updating behavior in important ways. At the population level, overinference and base-rate neglect are both significant in the baseline specification, but once the complete model is estimated, only base-rate neglect remains significant. At the individual level, the analysis uncovers substantial heterogeneity, and more subjects are categorized as exhibiting some type of bias. Comparing intermediate specifications further shows that overinference and preference-based biases are particularly sensitive to the inclusion of sequence effects. By contrast, the hot-hand fallacy remains comparatively stable when other biases are introduced and is the most commonly exhibited bias, while confirmatory distortions are comparatively scarce. Moreover, biases do not arise independently. Instead, they exhibit systematic co-occurrence patterns. Here, once again, the hot-hand fallacy remains central in the co-occurrence network, while biases like base-rate neglect also play a substantial role in connecting bias families.

Taken together, these findings suggest that studying belief-updating biases one at a time may give an incomplete picture of the underlying mechanisms shaping inference. More generally, a framework that separates multiple competing distortions can be useful for researchers interested in the interaction between biases, as well as for work seeking belief-based explanations of broader behavioral phenomena, such as political polarization, investment behavior, or the disposition effect.

Several directions for future research remain open. While this paper takes detailed care in separating behavior stemming from different biases, it treats such biases as systematic departures from Bayesian updating and does not attempt to identify the deeper cognitive mechanisms through which they arise. In this direction, \citet{bordalo2025people} study how some biases become more prominent when salient features of information become more relevant. Likewise, \citet{bordalo2023memory} emphasize the role of memory and recall in generating well-known cognitive distortions, while \citet{graeber2024stories} show how different forms of information --such as narratives versus statistical evidence-- affect belief updating. Exploring these mechanisms in settings where biases can already be separately identified may help provide a more complete account of which updating distortions matter most, and of the underlying primitives that generate biased behavior.

\newpage
\bibliography{References}
\bibliographystyle{apalike}




\appendix

\newpage

\centerline{\huge Appendix}

\section{Theoretical derivations}
\label{sec.Conjugate analyses}

This section shows the derivation of equations of posterior parameters in section 2. I present two cases: Proposition 1 characterizes the Bayesian benchmark under standard beta-binomial conjugacy (eq.(3)), and Proposition 2 introduces the biases of the baseline model (eq.(4) and (5)). The rest of the derivations for the remaining biases are analogous to these cases. Once the distorted likelihood and prior are given, the proof follows the exact same steps.

\begin{proposition}
Given likelihood equation (1) and prior equation (2) of Section 2.1., a Bayesian agent updates her beliefs such that her posterior distribution of $p$: $\pi(p|a_n, b_n)$  is beta distributed with parameters $a_n, b_n$ such that: 
$$
a_n = k + a_0 \qquad b_n = n-k + b_0  
$$

\end{proposition}

\begin{proof}

Let us apply Bayes' Theorem given likelihood eq.(1) and prior eq.(2). This yields:

\begin{align*}
    \pi(p|s_1...s_n, a_0 , b_0) &= \frac{L(p|s_1 ... s_n) \pi(p|a_0, b_0)}{\int_{p=0}^{1} \big(L(p|s_1 ... s_n) \pi(p|a_0, b_0) \big) dp} \\
    &= \frac{\binom{n}{k} p ^{k + a_0 -1} (1-p)^{n - k + b_0 -1}/B(a_0, b_0)}{\int_{p=0}^{1} \Big(\binom{n}{k} p ^{k + a_0 -1} (1-p)^{n - k + b_0 -1}/B(a_0, b_0)\Big) dp}\\
    &= \frac{p ^{k + a_0 -1} (1-p)^{n - k + b_0 -1}}{B(a_0 + k , b_0 + n - k)}
\end{align*}

Which is itself the probability density function of a beta distribution with parameters ($a_0 + k, b_0 + n - k $). Therefore, the posterior Bayesian distribution is beta distributed with parameters ($a_n,b_n$) as defined in Proposition 1.

\end{proof}

\begin{proposition}
    Given likelihood $\Tilde{L}(p|s_1 ... s_n) = (L(p|s_1 ... s_n))^\gamma$ and prior $\Tilde{\pi}(p) = (\pi(p))^\delta$ of section 2.1., where $\gamma, \delta$ indicate deviations from Bayesian updating, a non-Bayesian agent follows a posterior beta distribution with parameters $\Tilde{a}_n,\Tilde{b}_n$ such that:
\begin{equation*}
\label{eq.non-bayes S_a trial}
    \Tilde{a}_n = \gamma k + \delta (a_0 - 1) + 1
\end{equation*}
\begin{equation*}
\label{eq.non-bayes S_b trial}
    \Tilde{b}_n = \gamma (n -k) + \delta (b_0 - 1) + 1
\end{equation*}

\end{proposition}

\begin{proof}

Applying Bayes' theorem yields:

\begin{align*}
    \Tilde{\pi}(p|s_1...s_n, a_0 , b_0) &= \frac{(L(p|s_1 ... s_n))^\gamma (\pi(p))^\delta}{\int_{p=0}^{1} \big((L(p|s_1 ... s_n))^\gamma (\pi(p))^\delta \big) dp} \\
    &= \frac{\big[\binom{n}{k}\big]^{\gamma} p ^{\gamma k} (1-p)^{\gamma (n - k)}\big( 1/B(a_0, b_0)\big)^{\delta}p^{\delta(a_0 -1)}(1-p)^{\delta(b_0 -1)}}{\int_{p=0}^{1} \Big(\big[\binom{n}{k}\big]^{\gamma} p ^{\gamma k} (1-p)^{\gamma (n - k)}\big( 1/B(a_0, b_0)\big)^{\delta}p^{\delta(a_0 -1)}(1-p)^{\delta(b_0 -1)}\Big) dp}\\
    &=\frac{p^{\gamma k + \delta(a_0 -1)} (1-p)^{\gamma(n-k) + \delta(b_0-1)}}{\int_{p=0}^{1}\Big(p^{\gamma k + \delta(a_0 -1)} (1-p)^{\gamma(n-k) + \delta(b_0-1)}\Big)dp}\\
    &=\frac{p^{\gamma k + \delta(a_0 -1)} (1-p)^{\gamma(n-k) + \delta(b_0-1)}}{B\Big(\gamma k + \delta(a_0 -1) +1,\gamma(n-k) + \delta(b_0-1) +1 \Big)}
\end{align*}

This is the probability density function of a beta distribution with parameters $\Big(\gamma k + \delta(a_0 -1) +1,\gamma(n-k) + \delta(b_0-1) +1\Big)$. Therefore, the posterior distribution of an agent that exhibits inference bias $\gamma$, and base-rate bias ($\delta$) is beta distributed with parameters ($\Tilde{a}_n,\Tilde{b}_n$) as defined in Proposition 2.
\end{proof}

\section{Incentivizing moments of beta distributions}
\label{sec.scoring rules}

The scoring rules follow \citet{schlag2013eliciting} with a slight modification. In particular, random realizations of the Bayesian posterior distributions are taken as the random draws. 

\paragraph{Incentivizing the mean.}Let $\Tilde{m}$ be the reported mean of the agent's posterior beta distribution $\pi(p|\Tilde{a}_n,\Tilde{b}_n)$, and let $d$ be a random draw of the Bayesian posterior beta distribution $\pi(p|a_n,b_n)$. Then the Quadratic Scoring Rule is given by $g_{QSR}(\Tilde{m},d)=-(\Tilde{m}-d)^2$. Let $A,B$ be the boundaries of the state-space $\Omega$, and $M$ be any arbitrary amount of money.\footnote{In the experiment $A=1$, $B=99$ and $M=25/3$ cents for each report (that is a total maximum of 10 euros)} Then, the randomized quadratic scoring rule is given by the following lottery:
\begin{equation*}
\Tilde{g}_{QSR}(\Tilde{m},d)=l \Bigg(M,0;1+\frac{g_{QSR}(\Tilde{m},d)}{(B-A)^2}\Bigg)    
\end{equation*}

\paragraph{Incentivizing the variance.}Let $\Tilde{v}$ be the reported variance of the agent's posterior beta distribution. In order to elicit the variance consider two independent random draws of the Bayesian posterior beta distribution $\pi(p|a_n,b_n)$. Namely, $d_1$ and $d_2$. Then the variance scoring rule is given by $g_{v}(\Tilde{v},d_1,d_2)=-\Big(\Tilde{v}-\frac{1}{2}(d_1-d_2)^2\Big)^2$. Applying randomization, the randomized variance scoring rule yields:
\begin{equation*}
    \Tilde{g}_{v}(\Tilde{v},d_1,d_2)=l \Bigg(M,0;\frac{g_{v}+\frac{1}{4}(B-A)^4}{\frac{1}{4}(B-A)^4}\Bigg)    
\end{equation*}

\newpage

\section{Extra tables and figures}
\label{sec.extra}

\begin{table}[htbp]
    \centering\scriptsize
    \begin{tabular}{lccc}
        \toprule
        \textbf{Model} & \textbf{$R^2$ Population} & \textbf{Mean $R^2$ Ind.} & \textbf{Mean adj.\ $R^2$ Ind.} \\
        \midrule
        \textbf{Baseline model} & & & \\
        \hspace{0.5cm} Eq. (12) & 0.002 & 0.565 & 0.534 \\
        \hspace{0.5cm} Eq. (13) & 0.003 & 0.535 & 0.501 \\
        \midrule
        \textbf{Complete model} & & & \\
        \hspace{0.5cm} Eq. (14) & 0.003 & 0.663 & 0.596 \\
        \hspace{0.5cm} Eq. (15) & 0.004 & 0.714 & 0.656 \\
        \hspace{0.5cm} Eq. (16) & 0.292 & 0.321 & 0.297 \\
        \bottomrule
    \end{tabular}
        \caption{$R^2$ Comparison for Baseline and Complete model at population vs.\ individual level}
            \label{tab:R2 comp}
\end{table}

\begin{table}[htbp]
    \centering\scriptsize
    \begin{tabular}{lcccc}
        \toprule
        \textbf{Model} & \textbf{AIC(pop.)} & \textbf{BIC(pop.)}
        & \textbf{AIC(ind.)} & \textbf{BIC(ind.)} \\
        \midrule
        \textbf{Baseline model} & & & & \\
        \hspace{0.5cm} Eq. (12)
        & 49858.50 & 49876.13 & 23708.96 & 25260.39 \\
        \hspace{0.5cm} Eq. (13)
        & 51038.89 & 51056.52 & 23326.16 & 24877.59 \\
        \midrule
        \textbf{Complete model} & & & & \\
        \hspace{0.5cm} Eq. (14)
        & 49859.49 & 49894.75 & 23527.68 & 26630.55 \\
        \hspace{0.5cm} Eq. (15)
        & 51041.91 & 51077.17 & 22440.00 & 25542.87 \\
        \hspace{0.5cm} Eq. (16)
        & -14672.77 & -14655.14 & -16456.88 & -14905.45 \\
        \bottomrule
    \end{tabular}
    \caption{Information criteria for pooled and individual estimation. Individual criteria are calculated jointly across participants,
accounting for all participant-specific parameters.}
    \label{tab:info crit}
\end{table}

\begin{figure}[!htbp]
    \centering
    \captionsetup{
        font=scriptsize,
        skip=4pt,
        justification=justified,
        singlelinecheck=false
    }

    \includegraphics[
        width=0.95\textwidth,
        height=0.35\textheight,
        keepaspectratio
    ]{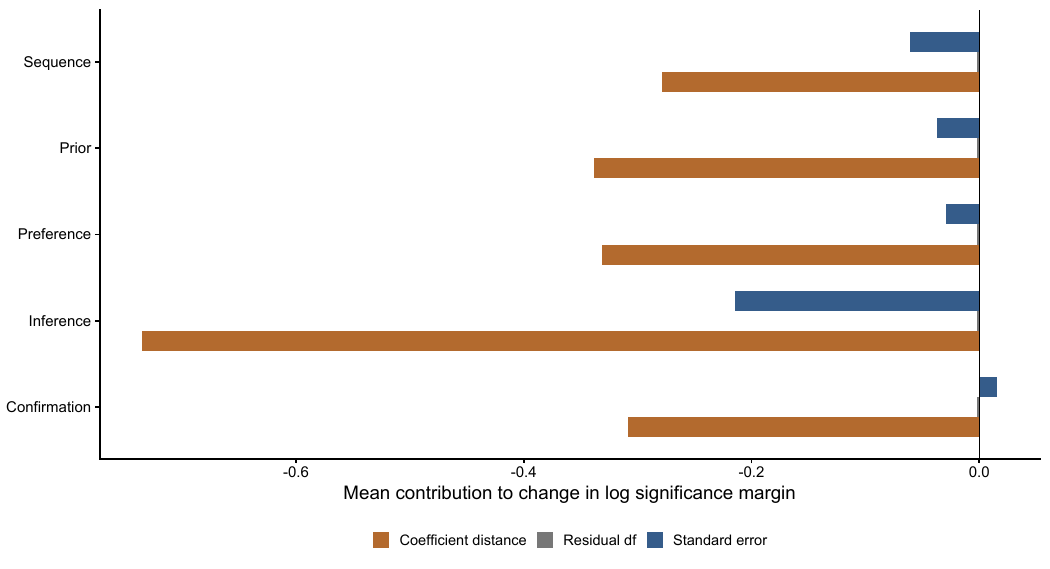}

    \caption{Decomposition of losses of statistical significance.
    Bars report mean contributions of coefficient distance from
    the Bayesian benchmark, standard errors and residual degrees
    of freedom to changes in the log significance margin, among
    tests that become insignificant.}
    \label{fig:misattribution_coefficient_se}
\end{figure}

\begin{figure}[!htbp]
\centering
\captionsetup{
    font=scriptsize,
    skip=4pt,
    justification=justified,
    singlelinecheck=false
}

\begin{minipage}[t]{0.54\textwidth}
    \vspace{0pt}
    \centering
    \includegraphics[width=\linewidth]{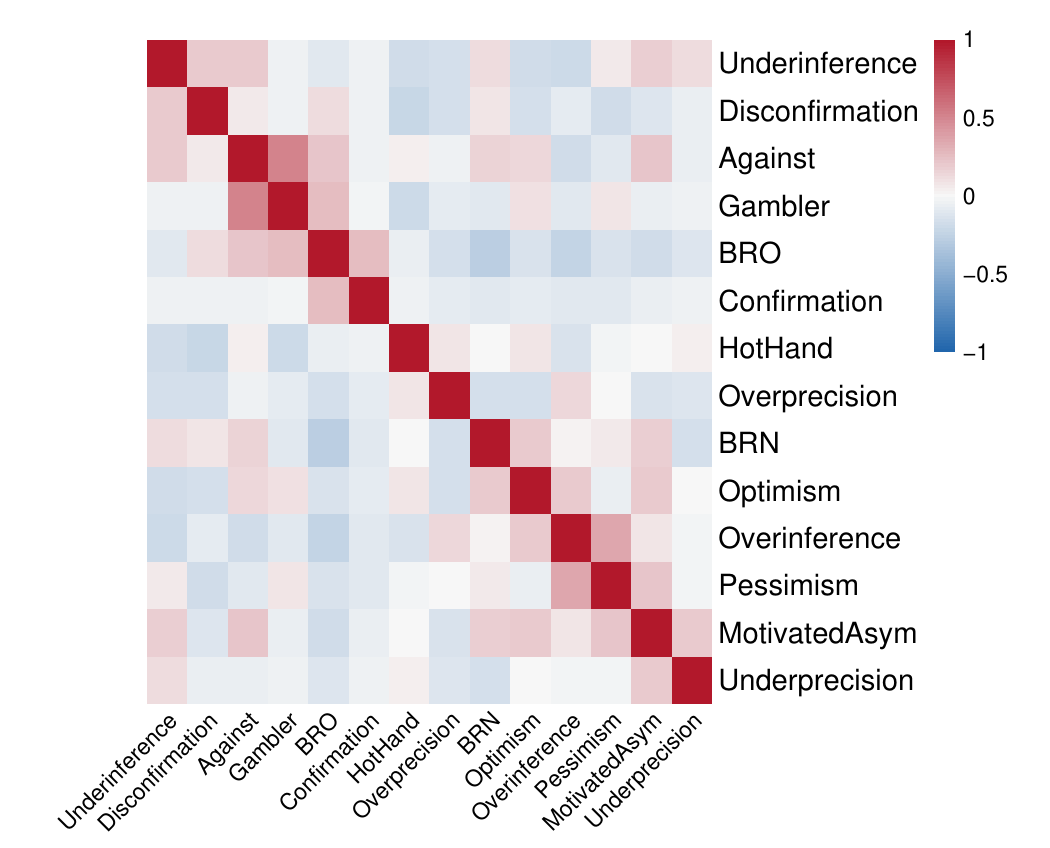}

    \captionof{figure}{Bias co-occurrence heatmap. Cells report
    pairwise $\phi$ correlations between bias indicators.}
    \label{fig:cooccur_heatmap}
\end{minipage}
\hfill
\begin{minipage}[t]{0.45\textwidth}
    \vspace{0pt}
    \centering
    \scriptsize
    \renewcommand{\arraystretch}{1.0}
    \setlength{\tabcolsep}{3pt}

    \begin{tabular}{@{}p{0.46\linewidth}rrr@{}}
        \toprule
        \textbf{Bias} & \textbf{Degree} &
        \textbf{Strength} & \textbf{Betweenness} \\
        \midrule
        Hot-hand fallacy        & 10 & 0.678 & 0.564 \\
        Base-rate neglect      &  4 & 0.460 & 0.135 \\
        Pessimism              &  3 & 0.568 & 0.006 \\
        Against the signal
                               &  3 & 0.419 & 0.013 \\
        Motivated asymmetry    &  3 & 0.382 & 0.013 \\
        Overinference          &  2 & 0.486 & 0.000 \\
        Base-rate overuse      &  2 & 0.287 & 0.000 \\
        Gambler's fallacy      &  1 & 0.207 & 0.000 \\
        Underinference         &  1 & 0.123 & 0.000 \\
        Overprecision          &  1 & 0.079 & 0.000 \\
        Optimism               &  1 & 0.078 & 0.000 \\
        Underprecision         &  1 & 0.047 & 0.000 \\
        Confirmation           &  0 & 0.000 & 0.000 \\
        Disconfirmation        &  0 & 0.000 & 0.000 \\
        \bottomrule
    \end{tabular}

    \captionof{table}{Connectivity and centrality
    in the bias co-occurrence network.}
    \label{tab:network_centrality}

    \par\smallskip
    \raggedright
    \textit{Notes:} Degree counts incident links; strength sums
    their absolute $\phi$ correlations. Betweenness is normalized
    to $[0,1]$ and calculated from shortest paths in the unweighted
    network. All measures use the edges in
    Figure~\ref{fig:cooccur_network}.
\end{minipage}

\par\vspace{1em}

\end{figure}

\newpage

\section{Robustness analyses}

\subsection{Robustness to Measurement Error in Uncertainty Reports}
\label{subsec.Measurement_Error}

Uncertainty is elicited through a continuous slider and may therefore be measured with classical measurement error. To assess whether such measurement error affects the bias classification, I borrow the noise-injection logic from the SIMEX (Simulation Extrapolation) method of \citet{Cook1994SIMEX} for parametric models, and apply it to both the prior and posterior reported variance.

Let $\mu$ denote the reported belief mean (be it prior or posterior) and $V$ the reported variance. Because beliefs are defined on the bounded support $[0,1]$, reported variances must satisfy
$$
0 < V < V_{\max}(\mu), \qquad 
V_{\max}(\mu)=\mu(1-\mu)
$$ 

Note that this restriction follows from the support of the distribution and does not impose unimodality of the implied beta distribution; bimodal beta distributions remain feasible whenever the reported variance is sufficiently close to the upper bound.
To model reporting noise while preserving feasibility, I work with the normalized variance ratio
$$
u=\frac{V}{V_{\max}(\mu)}\in(0,1)
$$
and apply the logit transformation $z=\log\!\left(\frac{u}{1-u}\right)\in\mathbb{R}$.

Measurement error is assumed to be classical and additive on this scale:
$$
z^{\text{rep}} = z^{\star} + \sigma_z \varepsilon
\qquad \varepsilon \sim \mathcal{N}(0,1)
$$
where $z^{\star}$ denotes the latent (noise-free) report and $\sigma_z$ is the standard deviation of reporting noise on the logit-normalized variance scale.

\begin{figure}[!htbp]
    \centering
    \captionsetup[subfigure]{font=scriptsize,justification=centering,
        singlelinecheck=false,skip=3pt}

    \begin{subfigure}[t]{0.48\textwidth}
        \centering
        \includegraphics[width=\linewidth,height=0.40\textheight,keepaspectratio]{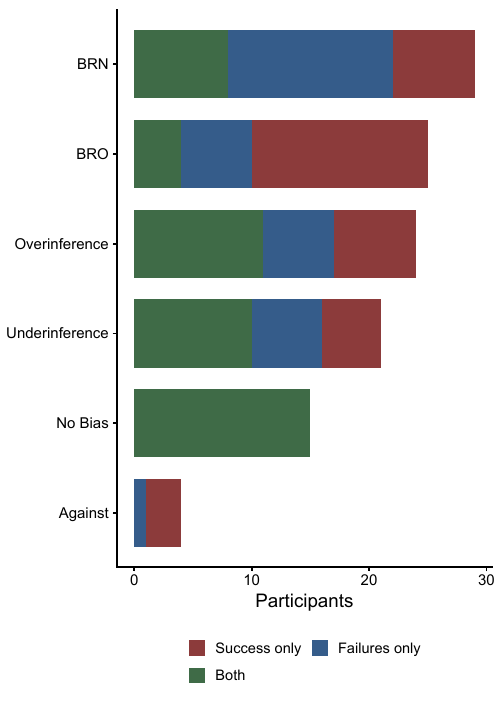}
        \caption{Baseline model.}
        \label{fig:noise_baseline}
    \end{subfigure}
    \hfill
    \begin{subfigure}[t]{0.48\textwidth}
        \centering
        \includegraphics[width=\linewidth,height=0.40\textheight,keepaspectratio]{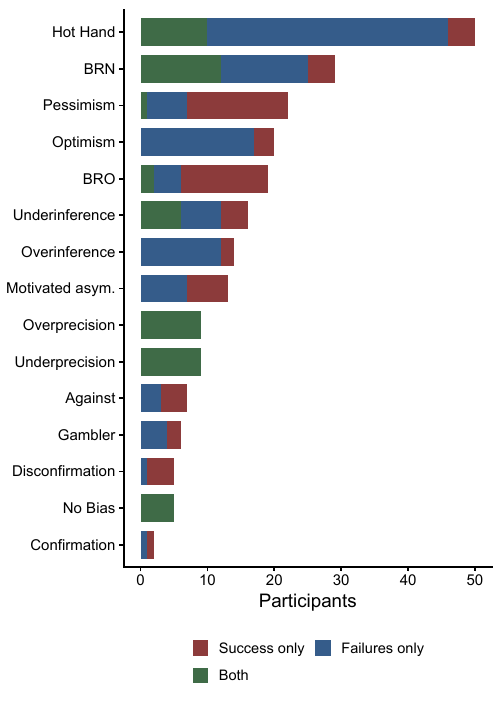}
        \caption{Complete model.}
        \label{fig:noise_complete}
    \end{subfigure}

    \par\vspace{0.6em}

    \begin{subfigure}[t]{0.43\textwidth}
        \centering
        \includegraphics[width=\linewidth,height=0.33\textheight,keepaspectratio]{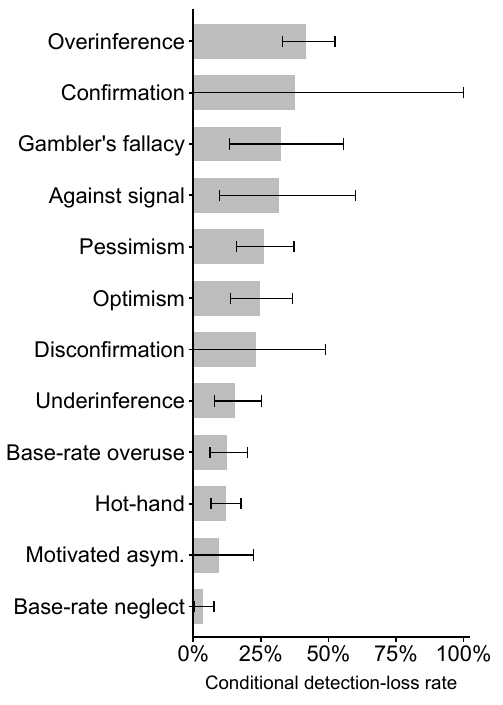}
        \caption{Bias misattribution sensitivity.}
        \label{fig:noise_sensitivity}
    \end{subfigure}
    \hfill
    \begin{subfigure}[t]{0.56\textwidth}
        \centering
        \includegraphics[width=\linewidth,height=0.33\textheight,keepaspectratio]{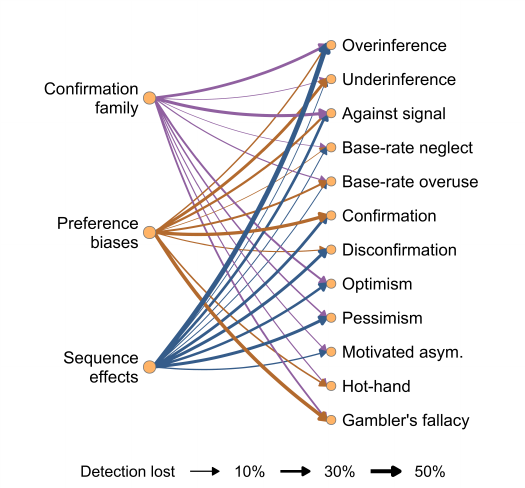}
        \caption{Detection losses by added bias family.}
        \label{fig:noise_misattribution_network}
    \end{subfigure}

    \caption{Noise injection ($\lambda=0.5$): bias prevalence and misattribution.}
    \label{fig:noise_prevalence_misattribution}
    \par\smallskip
    \begin{minipage}{0.97\textwidth}
        \footnotesize
        \textit{Notes:} All panels use $p<0.05$ after noise injection at $\lambda=0.5$.
        Panels (a)--(b) report the number of participants exhibiting each bias.
        Panel (c) reports the fraction of initially detected classifications
        lost when one additional bias family is included. Rates based on
        three or fewer participants contributing initially detected
        classifications are omitted from panel (c). Error bars are
        95\% percentile confidence intervals from 1,000 participant bootstrap
        samples, retaining each participant's full set of model comparisons.
        In panel (d), arrows run from the added family to the affected bias.
        Thickness indicates the corresponding conditional detection-loss rate.
    \end{minipage}
\end{figure}

\begin{figure}[!htbp]
    \centering
    \captionsetup[subfigure]{font=scriptsize,justification=centering,
        singlelinecheck=false,skip=3pt}

    \begin{subfigure}[t]{0.48\textwidth}
        \centering
        \includegraphics[width=\linewidth,height=0.33\textheight,keepaspectratio]{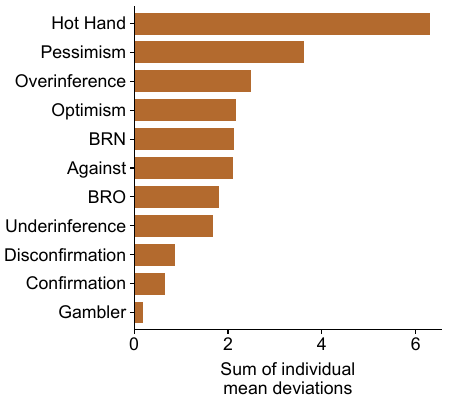}
        \caption{Expected-value deviations.}
        \label{fig:noise_ev_weighted}
    \end{subfigure}
    \hfill
    \begin{subfigure}[t]{0.48\textwidth}
        \centering
        \includegraphics[width=\linewidth,height=0.33\textheight,keepaspectratio]{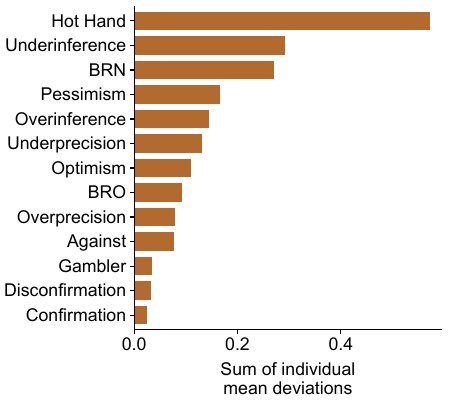}
        \caption{Variance deviations.}
        \label{fig:noise_variance_weighted}
    \end{subfigure}

    \caption{Noise injection ($\lambda=0.5$): prevalence-adjusted deviations from Bayesian updating.}
    \label{fig:noise_impacts}
    \par\smallskip
    \begin{minipage}{0.97\textwidth}
        \footnotesize
        \textit{Notes:} Each panel sums individual mean absolute deviations from the
        Bayesian benchmark across participants exhibiting the corresponding
        bias. Only the parameters associated with that bias depart from their
        Bayesian values in each exercise. Noise is injected at $\lambda=0.5$.
    \end{minipage}
\end{figure}

To implement this robustness exercise, I simulate additional noise on the logit-normalized variance scale. For each $\lambda\in\Lambda=\{0.1,0.25,0.5\}$ and each observation, the perturbed report is given by
\begin{equation}
    z^{(\lambda)} = z^{\text{rep}} + \sqrt{\lambda}\,\sigma_z\,\varepsilon^{(\lambda)}
\qquad \varepsilon^{(\lambda)}\sim\mathcal{N}(0,1)
\end{equation}
where $\lambda$ indexes the severity of the perturbation. Thus, $\lambda=0.1$, $\lambda=0.25$, and $\lambda=0.5$ correspond to adding noise with variance equal to $10\%$, $25\%$, and $50\%$ of the estimated baseline measurement-error variance, respectively. This range is intended to capture progressively stronger perturbations of the reported variance while remaining within a plausible range for slider misreporting.

A baseline value $\sigma_z=\hat\sigma_z$ is calibrated from the data as the residual standard deviation from a regression of reported logit-normalized variances on their Bayesian benchmark variances on the same scale. Mapping back via $u^{(\lambda)}=\text{logit}^{-1}(z^{(\lambda)})$ yields simulated feasible variances (both for prior and posterior measurements)
$$
V^{(\lambda)} = u^{(\lambda)} V_{\max}(\mu)
$$

For each $\lambda$, the simulation is repeated $200$ times and the adjusted variances $V^{(\lambda)}$ are averaged across repetitions. The procedure is applied to both prior and posterior variance reports. The implied beta-distribution parameters are recomputed and all model equations are re-estimated using these noise-adjusted values.

Figures \ref{fig:noise_prevalence_misattribution} and \ref{fig:noise_impacts} report the findings for the strongest perturbation, i.e., $\lambda=0.5$. The main qualitative patterns remain robust to this noise injection: the complete model categorizes substantially more participants than the baseline model, each bias category is identified for some participants, and the hot-hand fallacy remains the most frequently exhibited bias.

The misattribution exercise also retains the most important results. Overinference remains sensitive to the introduction of confounding biases, losing significance in roughly $42\%$ of initially detected comparisons. Optimism and pessimism also remain sensitive (loss rates are approximately $25\%$ and $26\%$, respectively), and just like in the main analysis, hot-hand fallacy and base-rate neglect remain the least sensitive to the inclusion of additional distortions with loss rates of approximately $12\%$, and $3.5\%$, respectively. The directed network also continues to indicate that incorporating sequence effects generates particularly substantial losses of overinference, and also decreases the detection of optimism and pessimism.

Finally, Figure \ref{fig:noise_impacts} shows that the hot-hand fallacy and pessimism remain important contributors to departures from Bayesian posterior expected values and variances. Importantly, underinference becomes comparatively more prominent in variance deviations after noise injection, but the general conclusions remain unchanged.

\subsection{Robustness to Multiple Hypothesis Testing}
\label{sec.robust}

Since the individual-level analysis in Section \ref{sec.ind analysis} categorizes subjects on the basis of multiple coefficient tests, a reasonable concern is that some identified biases may represent false positives. This issue is especially relevant when comparing the baseline and complete models, because the baseline model tests four hypotheses per participant, while the complete model tests twelve. To address this concern, I implement a Bonferroni correction within each model. Consequently, the significance cutoff is set to $p<0.05/4=0.0125$ for the baseline model and to $p<0.05/12\approx0.0042$ for the complete model.

\begin{figure}[!htbp]
    \centering
    \captionsetup[subfigure]{font=scriptsize,justification=centering,
        singlelinecheck=false,skip=3pt}

    \begin{subfigure}[t]{0.48\textwidth}
        \centering
        \includegraphics[width=\linewidth,height=0.40\textheight,keepaspectratio]{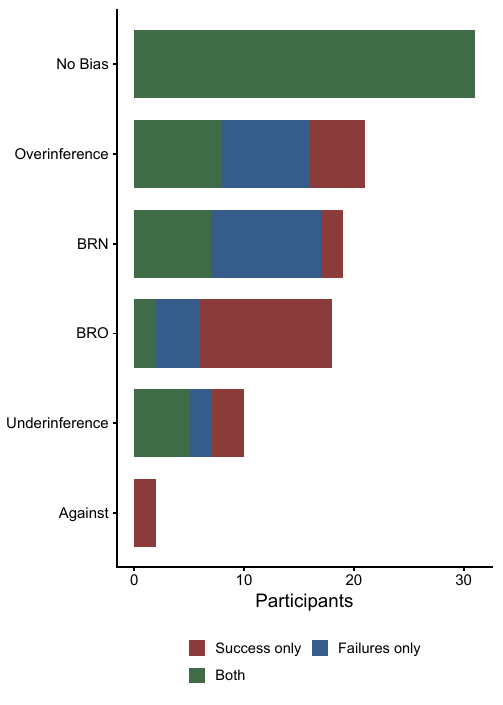}
        \caption{Baseline model.}
        \label{fig:mht_baseline}
    \end{subfigure}
    \hfill
    \begin{subfigure}[t]{0.48\textwidth}
        \centering
        \includegraphics[width=\linewidth,height=0.40\textheight,keepaspectratio]{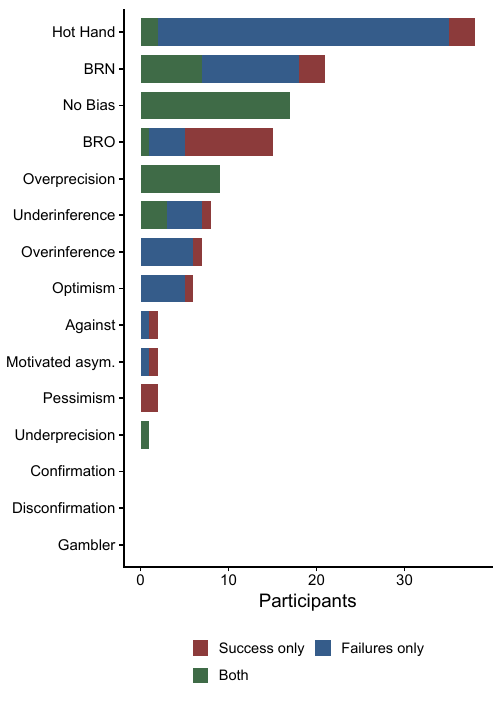}
        \caption{Complete model.}
        \label{fig:mht_complete}
    \end{subfigure}

    \par\vspace{0.6em}

    \begin{subfigure}[t]{0.43\textwidth}
        \centering
        \includegraphics[width=\linewidth,height=0.33\textheight,keepaspectratio]{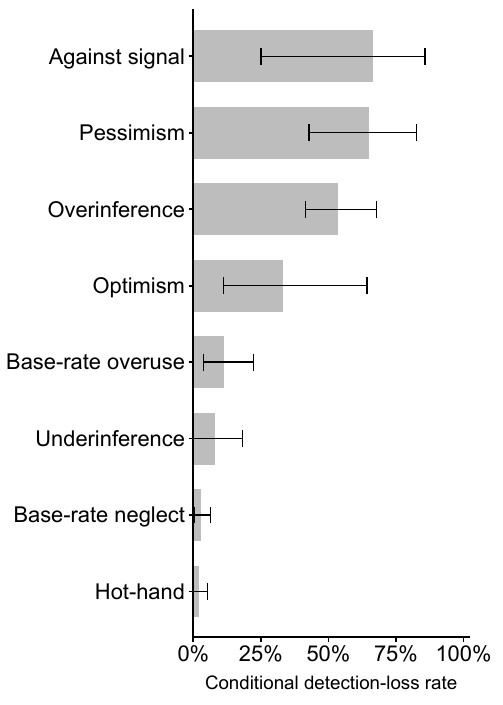}
        \caption{Bias misattribution sensitivity.}
        \label{fig:mht_sensitivity}
    \end{subfigure}
    \hfill
    \begin{subfigure}[t]{0.56\textwidth}
        \centering
        \includegraphics[width=\linewidth,height=0.33\textheight,keepaspectratio]{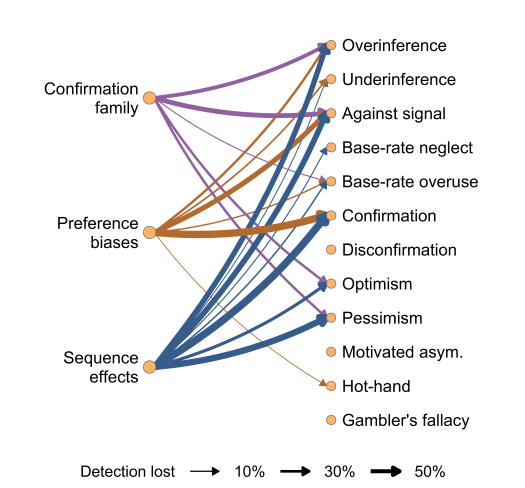}
        \caption{Detection losses by added bias family.}
        \label{fig:mht_misattribution_network}
    \end{subfigure}

    \caption{Multiple-hypothesis correction: bias prevalence and misattribution.}
    \label{fig:mht_prevalence_misattribution}
    \par\smallskip
    \begin{minipage}{0.97\textwidth}
        \footnotesize
        \textit{Notes:} Panel (a) uses $p<0.05/4$; panel (b) uses $p<0.05/12 \approx 0.0041$. Panels (c)--(d) hold $p< 0.05/12$ fixed across all eight specifications.
        Panels (a)--(b) report the number of participants exhibiting each bias.
        Panel (c) reports the fraction of initially detected classifications
        lost when one additional bias family is included. Rates based on
        three or fewer distinct participants contributing initially detected
        classifications are omitted from panel (c) only. Error bars are
        95\% percentile confidence intervals from 1,000 participant bootstrap
        samples, retaining each participant's full set of model comparisons.
        In panel (d), arrows run from the added family to the affected bias;
        thickness indicates the corresponding conditional detection-loss rate.
    \end{minipage}
\end{figure}

\begin{figure}[!htbp]
    \centering
    \captionsetup[subfigure]{font=scriptsize,justification=centering,
        singlelinecheck=false,skip=3pt}

    \begin{subfigure}[t]{0.48\textwidth}
        \centering
        \includegraphics[width=\linewidth,height=0.33\textheight,keepaspectratio]{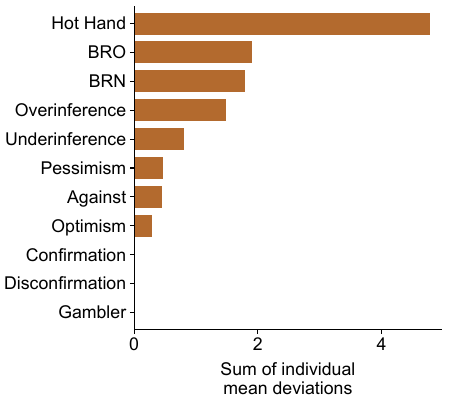}
        \caption{Expected-value deviations.}
        \label{fig:mht_ev_weighted}
    \end{subfigure}
    \hfill
    \begin{subfigure}[t]{0.48\textwidth}
        \centering
        \includegraphics[width=\linewidth,height=0.33\textheight,keepaspectratio]{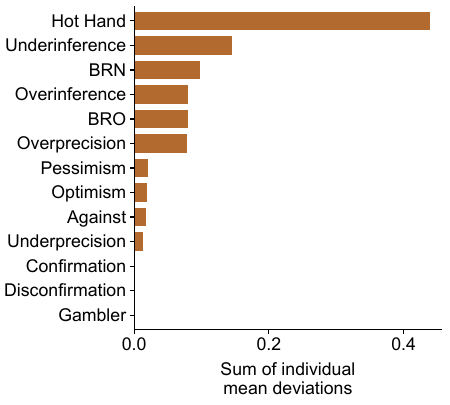}
        \caption{Variance deviations.}
        \label{fig:mht_variance_weighted}
    \end{subfigure}

    \caption{Multiple-hypothesis correction: prevalence-adjusted deviations from Bayesian updating.}
    \label{fig:mht_impacts}
    \par\smallskip
    \begin{minipage}{0.97\textwidth}
        \footnotesize
        \textit{Notes:} Each panel sums individual mean absolute deviations from the
        Bayesian benchmark across participants exhibiting the corresponding
        bias. Only the parameters associated with that bias depart from their
        Bayesian values in each exercise. Detection uses the complete-model cutoff $p<0.05/12$.
    \end{minipage}
\end{figure}

For the misattribution exercise, I apply the same threshold $p<0.05/12$ across all eight specifications, to ensure that a bias classification cannot disappear simply because introducing a new family changes the significance threshold. Figures \ref{fig:mht_prevalence_misattribution} and \ref{fig:mht_impacts} present the adjusted results.

The main qualitative conclusions remain largely unchanged: There is substantial individual heterogeneity across all bias categories, the hot-hand fallacy continues to be the most commonly exhibited bias and the prevalence of overinference diminishes in the complete model (See Figures \ref{fig:mht_baseline} and \ref{fig:mht_complete}). Also, the hot-hand fallacy continues to play a leading role in both expected-value and variance deviations (See Figures \ref{fig:mht_ev_weighted} and \ref{fig:mht_variance_weighted}). Consistent with the rest of the results of the paper, the confirmation-bias family disappears under a more demanding threshold. The misattribution results are also robust: overinference and preference-based biases remain sensitive to competing explanations, while hot-hand fallacy and base-rate neglect remain comparatively stable (Figure \ref{fig:mht_sensitivity}). The directed network, once again, shows that introducing sequence effects reduces the detection of overinference, optimism and pessimism (Figure \ref{fig:mht_misattribution_network}).

At the same time, the correction naturally reduces the number of detected significant biases and increases the number of unclassified subjects in both models. Nevertheless, the complete model still classifies substantially more subjects than the baseline model. In particular, the number of subjects in the ``No Bias'' category falls from $31$ under the baseline model to $17$ under the complete model (a $\approx 45\%$ reduction). Finally, it is important to note that preference-based biases --especially optimism and pessimism-- become relatively less prevalent under the corrected thresholds.


\section{Estimating a Grether-style reduced-form model}
\label{sec.Grether_Check}

This section connects the main continuous state-space analysis to the classic reduced-form literature on belief updating in the spirit of \citet{grether1980bayes}. To do so, I approximate the continuous state space of the model by a binary state space and estimate a Grether-style regression using the experimental data. This exercise is intended to show that applying the usual models to this experimental data recovers patterns that can be related both to the classic literature and to the baseline findings of the main model.

Let the state space $\bar{\Omega}=\{L,H\}$ be a binary partition of the original state-space $\Omega=(0,1)$, where state $H$ corresponds to higher success probabilities and state $L$ to lower success probabilities.

Let $q_0=\Pr(H)$ denote the prior belief assigned to the high state, and let $q_n=\Pr(H\mid k,n)$ denote the corresponding posterior belief after observing $k$ successes in $n$ Bernoulli trials. Conditional on state $\bar{\omega}\in\{L,H\}$, the probability of observing $k$ successes is
$$
\Pr(k\mid \bar{\omega})
=
\int_{p\in (0,1)}
\binom{n}{k}p^k(1-p)^{n-k}f(p\mid \bar{\omega})\,dp
$$
where $f(p\mid \bar{\omega})$ denotes the prior density over $p$ conditional on state $\bar{\omega}$.

\noindent The associated log-likelihood ratio is therefore
$$
\ell(k,n)
=\log\left(\frac{\Pr(k\mid H)}{\Pr(k\mid L)}\right)
$$
Under Bayesian updating, posterior log-odds satisfy
$$
\log\left(\frac{q_n}{1-q_n}\right)
=
\log\left(\frac{q_0}{1-q_0}\right)
+\ell(k,n)
$$

A Grether-style reduced form allows the prior and the likelihood ratio to receive separate weights in the updating equation. I therefore estimate
\begin{equation}
\log\left(\frac{\tilde q_n}{1-\tilde q_n}\right)
=
\beta_0
+\beta_P\log\left(\frac{q_0}{1-q_0}\right)
+\beta_L\,\ell(k,n)
+\varepsilon
\label{eq.grether}
\end{equation}
where $\tilde q_n$ denotes the reported posterior probability of the high state. The Bayesian benchmark is given by
$$
\beta_0=0 \qquad \beta_P=1 \qquad \beta_L=1
$$
Values $\beta_P<1$ indicate base-rate neglect, while $\beta_P>1$ indicate base-rate overuse. Similarly, $\beta_L<1$ indicates underinference, $\beta_L>1$ indicates overinference, and $\beta_L<0$ corresponds to inference against the signal, as in classic \citet{grether1980bayes}.

To implement this reduced-form exercise, I binarize the continuous state space at $p=0.5$, so that the high state is $H=\{p\geq 0.5\}$ and the low state is $L=\{p<0.5\}$. For each elicited prior and posterior beta distribution, I compute the implied probability assigned to the high state,
$q=\Pr(p\geq 0.5),$
and use these probabilities as the belief objects entering the Grether-style regression. The evidence term is computed as the Bayesian log-odds update implied by the observed signal under the same binary partition,
$
\log\!\left(\frac{q_n^B}{1-q_n^B}\right)-\log\!\left(\frac{q_0}{1-q_0}\right),
$
which is equivalent to $\ell(k,n)$ under this partition, where $q_0$ is the prior probability of the high state and $q_n^B$ is the Bayesian posterior probability of that state.

\begin{figure}[!htbp]
    \centering
    \begin{minipage}[t]{0.55\textwidth}
        \vspace{0pt}
        \centering
        \scriptsize
        \begin{tabular}{@{\extracolsep{5pt}}lc}
            \\[-1.8ex]\hline
            \hline \\[-1.8ex]
             & \textit{Dependent variable:} \\
            \cline{2-2}
            \\[-1.8ex] & Log-odds reported posterior \\
            \hline \\[-1.8ex]
            Prior log-odds & 0.681$^{***}$ \\
             & (0.047) \\
             & \\
            Bayesian log-likelihood ratio & 2.196$^{***}$ \\
             & (0.123) \\
             & \\
            Constant & 0.303$^{***}$ \\
             & (0.097) \\
             & \\
            \hline \\[-1.8ex]
            Observations & 2,635 \\
            R$^{2}$ & 0.559 \\
            Adjusted R$^{2}$ & 0.559 \\
            \hline
            \hline \\[-1.8ex]
            \textit{Note:} &
            $^{*}$p$<$0.1; $^{**}$p$<$0.05; $^{***}$p$<$0.01 \\
        \end{tabular}
        \captionof{table}{Grether-style model at the population level.
        Significance is with respect to Bayesian values.
        Clustered standard errors by participant.}
        \label{tab:grether}
    \end{minipage}
    \hfill
\begin{minipage}[t]{0.35\textwidth}
    \vspace{0pt}
    \includegraphics[width=0.7\linewidth]{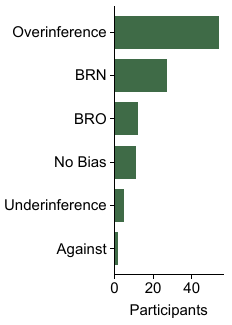}
    \caption{Number of times a bias is found to be significant
    ($p<0.05$) in the Grether-style model at the individual level.}
    \label{fig:grether}
\end{minipage}
\end{figure}

The Grether-style reduced-form estimates in Table \ref{tab:grether} are first and foremost consistent with the baseline analysis of the paper. At the population level, the prior-weight coefficient lies below the Bayesian benchmark of one ($\hat\beta_P=0.681$), indicating base-rate neglect, while the evidence-weight coefficient lies above one ($\hat\beta_L=2.196$), indicating overinference on average. Thus, once the data are forced into a traditional binary-state reduced form, the main qualitative pattern resembles that of the baseline model: subjects appear to underweight prior information and overweight new evidence.

At the same time, these estimates remain meaningfully connected to the classic reduced-form literature on belief updating. Base-rate neglect is the standard qualitative finding in the Grether tradition, and the presence of overinference, while less canonical than underinference \citep{benjamin2019errors}, is by no means implausible. More recent contributions also document environments in which subjects overweight new signals rather than underreact to them \citep{augenblick2025overinference, goncalves2026retractions}. In this sense, the Grether-style exercise places the data in a language that is comparable to the traditional literature, while still recovering the same broad reduced-form distortions highlighted by the baseline model.

The individual-level estimates (see Figure \ref{fig:grether}) likewise show substantial heterogeneity, again in a way that is closely aligned with the baseline specification. Overinference is the most prevalent reduced-form pattern, followed by base-rate neglect, while base-rate overuse, underinference and updating against the signal are also present for a small subset of participants.

\section{Identification with belief distributions}
\label{app:identification_gain}

This section formalizes the identification gain from observing belief distributions compared to projecting them onto a binary-state space. The argument is that insufficiently rich binary-state data will fail to identify multiple updating distortions. Therefore, using (beta) belief distributions, reduces the amount of updating environments (i.e. treatment variations) required for separate identification.

Let $\theta\in\Theta\subseteq\mathbb{R}^K$ be a vector of different updating parameters and let $t=1,\ldots,T$ denote distinct updating environments (that is, each belief-updating decision. An updating environment could be, for example, observing a specific number of blue and red balls.) Under distributional elicitation, we have
$$
m_t^D(\theta)
=
\begin{pmatrix}
a_t(\theta)\\
b_t(\theta)
\end{pmatrix},
$$

\noindent and let $M^D(\theta)$ stack these reports across environments.

Now, consider instead a binary-state representation obtained by applying a differentiable scalar projection $\psi:\mathbb{R}^2\rightarrow\mathbb{R}$ to
each distributional report:
$$
m_t^B(\theta)=\psi\left(m_t^D(\theta)\right),
$$

\noindent where $m^{B}_{t}(\theta)$ is the scalar projection of the Grether-style exercise in the previous subsection (i.e. equation \eqref{eq.grether}) , $\psi(a,b)$ is the log-odds transformation of $\Pr_{\mathrm{Beta}(a,b)}(p\geq 1/2)$.

Then suppose the researcher observes $T$ distinct updating environments, and wants to identify $K$ different bias parameters. Further, let $T_D(\theta)$ and $T_B(\theta)$ denote the minimum number of distinct updating environments required to locally identify $\theta$ from distributional and binary reports, respectively. Finally, let $J^D(\theta)$ and $J^B(\theta)$ denote the Jacobians of the distributional and binary report mappings with respect to $\theta$. 

The following proposition establishes the tradeoff between updating environments and distributional reports, and gives a natural sufficient condition (case (ii) of the proposition) for a situation where local identification is possible for beta-distributed reports, but fails in the case of binary reported beliefs. 

\begin{proposition}
\label{prop:distributional_identification}
\leavevmode\\
(i) Distributional reports weakly reduce the number of updating environments required for regular local identification. That is,
$T_D\leq T_B$.

\noindent
(ii) If $T<K\leq2T$ and $\operatorname{rank}J^D(\theta)=K$, then the above inequality is strict (i.e., $T_D<T_B$).
\end{proposition}

\begin{proof}
For any given set of updating environments, the scalar reports are differentiable projections of the distributional reports. By the chain rule, $J^B(\theta)=A(\theta)J^D(\theta)$, where $A(\theta)$ stacks the gradients of the scalar projection $\psi$. Therefore, $\operatorname{rank}J^B(\theta)\leq\operatorname{rank}J^D(\theta)$.

Hence, whenever a given set of updating environments regularly locally identifies $\theta$ from the scalar reports, so that $\operatorname{rank}J^B(\theta)=K$, the same set also satisfies $\operatorname{rank}J^D(\theta)=K$. Taking the minimum over sets of updating environments gives $T_D\leq T_B$, which proves (i).

For (ii), $\operatorname{rank}J^D(\theta)=K$ implies that the $T$ distributional reports regularly locally identify $\theta$, and therefore $T_D\leq T$. By contrast, the Jacobian associated with any $T$ scalar reports has $T$ rows, so $\operatorname{rank}J^B(\theta)\leq T<K$. This implies that there is no set of $T$ scalar updating environments which can satisfy the full-column-rank condition for regular local identification, so $T_B>T$. Combining the two inequalities yields $T_D\leq T<T_B$, and therefore $T_D<T_B$.
\end{proof}

\paragraph{Example: overinference and confirmation bias.}

Consider two distortions that may generate similar observed updating:
a general distortion in the weight placed on new information (i.e. overinference) and
confirmation bias, as defined in section \ref{sec.Theory}. To isolate the identification argument, impose symmetric
reactions to successes and failures and maintain Bayesian weighting of the
prior. Equations (9)--(10) then reduce to
\begin{align}
\tilde a_t-a_{0t}
    &= \gamma k_t+\rho c_t
    \label{eq:ident_example_a}\\
\tilde b_t-b_{0t}
    &= \gamma (n_t-k_t)+\rho c_t
    \label{eq:ident_example_b}
\end{align}
where $\gamma$ captures general over- or underinference and $\rho$ captures
the response to how confirming the signal is.

Suppose first that the researcher observes the full posterior distribution. For a single updating environment, the Jacobian of the two reported posterior parameters with respect to $(\gamma,\rho)$ is
$$
J_t^D
=
\begin{pmatrix}
k_t & c_t\\
n_t-k_t & c_t
\end{pmatrix}
$$
Its determinant is
$$
\det(J_t^D)
=
c_t\left(2k_t-n_t\right)
$$
Hence, whenever $c_t\neq0$ and the signal contains unequal numbers of successes and failures, the two distortions are locally separately identified from a single distributional belief report.

The intuition is immediate. For example, if a signal contains four successes and one failure,
$$
\tilde a_t-a_{0t}=4\gamma+\rho c_t
\qquad
\tilde b_t-b_{0t}=\gamma+\rho c_t
$$
Taking the difference identifies the general response to information,
$$
\gamma
=
\frac{(\tilde a_t-a_{0t})-(\tilde b_t-b_{0t})}{3}
$$
after which the confirmation component can be recovered from either equation. Thus, the two biases generate different movements across the two margins of the reported belief distribution.

Now suppose that the same posterior distribution is reduced to a scalar binary-state belief,
$$
m_t^B=\psi(\tilde a_t,\tilde b_t)
$$
The same updating environment then provides only one observable restriction
on $(\gamma,\rho)$. Its Jacobian is
$$
J_t^B
=
\begin{pmatrix}
\psi_a k_t+\psi_b(n_t-k_t)
&
(\psi_a+\psi_b)c_t
\end{pmatrix}
$$
which has rank at most one. A strong movement in the scalar posterior after a confirming signal can therefore be rationalized either by a high general weight on information or by a confirmation-specific distortion.

Separating these two biases with scalar belief reports requires richer data: the researcher must observe additional updating environments in which signal informativeness and confirmation vary sufficiently independently to provide a second restriction. In a conventional implementation in which confirmation is represented by a confirming-signal indicator (e.g. see \citet{benjamin2019errors}), this requires crossing confirming and disconfirming signals with sufficiently rich variation in signal strength or composition. If, for example, confirming signals are also systematically the stronger signals in the available data, general overinference and confirmation bias will not be separately recovered.

The bottom line is that distributional elicitation does not remove the need for experimental variation, but reduces this need in general, as it obtains two independent restrictions from the same updating decision. The richer belief object therefore substitutes, in part, for variation that would otherwise have to be generated across updating environments.

\section{Experiment instructions}
\label{sec. instructions}

In this study, you will be asked to complete \textbf{30 guessing tasks}. For each guessing task you have to make \textbf{2 related guesses}.
At the beginning of each guessing task, there is \textbf{always a pool of 99 URNS}, each containing \textbf{100 BALLS}.  Some balls in the urns are \textcolor{red}{\textbf{red}}, and some are \textcolor{blue}{\textbf{blue}}. Each one of these urns contains \textbf{a different percentage of \textcolor{red}{red} balls}. For example, in Urn 1 there is only one red ball and 99 blue (1\% of the balls are red), in Urn 2 there are only two red balls and 98 blue (2\% of the balls are red). This continues until Urn 99 where 99 balls are red and one is blue. (See picture below).

\begin{figure}[!htbp]
    \centering
    \includegraphics[width=11cm]{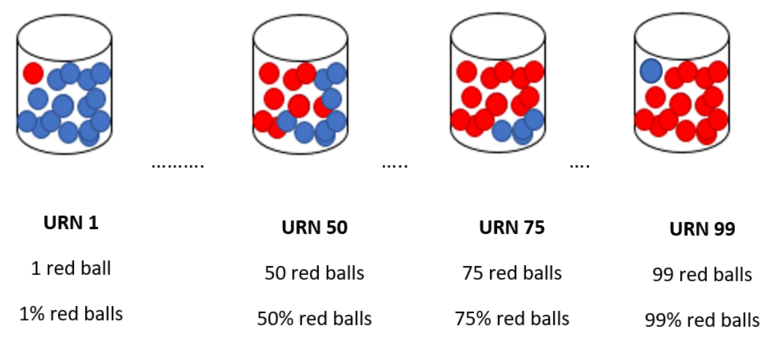}
    \label{fig:The Urns}
\end{figure}

For each guessing task, out of these 99 urns, one of them (say Urn X) has been selected \textbf{at random}.  Each urn has \textbf{the same chances} of being selected from the pool. That is, you do not know how many of the balls are red and how many of them are blue in the selected urn. All combinations are possible. (See picture below).

\begin{figure}[!htbp]
    \centering
    \includegraphics[width=11cm]{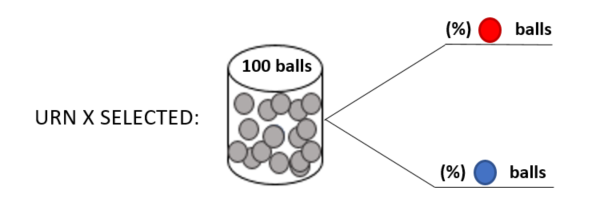}
    \label{fig:Urn proportion}
\end{figure}

Your task is to guess what the percentage of red balls in the selected urn is.
In each one of the 30 guessing tasks, this process will be \textbf{repeated with a new urn.}

Bear in mind: Whenever a new urn is selected, it is always drawn from the same pool of 99 urns (WITH REPLACEMENT). This means that the same urn can be chosen either once or multiple times. The urns have letters (or combinations of letters) on top of them. This is just to highlight the fact that when a new letter(or combination of letters) is on top of the urn, it means that a new urn has been drawn.

\vspace{2mm}

{\large \textbf{The guessing task}}

\vspace{2mm}

\noindent This section will explain how \textbf{each one} of the 30 guessing tasks works. Initially, as explained above, the computer has \textbf{randomly} selected, with equal probability, ONE out of the 99 urns. Remember that initially, you know nothing about the content of this urn. Once a given urn is selected, you will be given some information about the urn to help you make your guess.
First, you will see a sequence of balls which have been \textbf{randomly drawn} from the urn. Each ball in the selected urn has the same chances of being drawn, and each of these draws is done \textbf{WITH REPLACEMEN}T. This means that after a ball has been drawn and taken out of the urn, it is immediately replaced with one of the same color. \textbf{The urn will always have the same 100 balls}. In order to help you make your guesses, \textbf{two sequences of draws} will be made from each urn.

\begin{itemize}

\item \textbf{1st sequence of draws}: Either \textbf{one, two or three} balls from the urn are selected at first. After this selection, you will have to make \textbf{your first guess}. For your first guess you will answer \textbf{two questions}: 

\begin{enumerate}
    \item \textbf{\textit{What percentage of \textcolor{red}{red balls} do you expect the selected urn to have?}}
    \item \textbf{\textit{What is your uncertainty level about this percentage?}}
\end{enumerate}

\textit{Example: Guessing Task A}

\vspace{2mm}

\includegraphics[width=14cm]{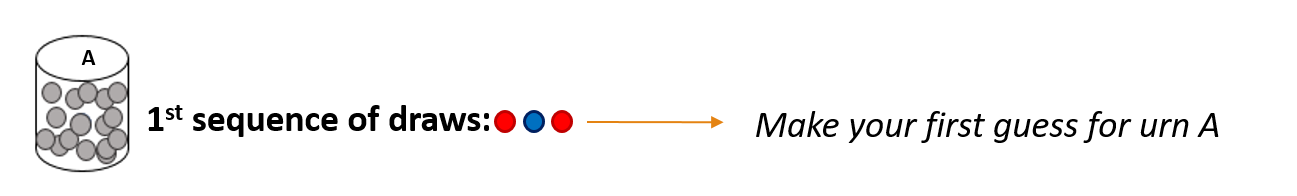}

\item \textbf{2nd sequence of draws}: The second draw follows very similar rules to the first one. In this case \textbf{three, five or seven balls} from \textbf{THE SAME URN} are selected with replacement. After this selection, you will have to make your \textbf{second guess}. Once again, for your second guess you will answer the same \textbf{two questions}:
 
\begin{enumerate}
    \item \textbf{\textit{What percentage of \textcolor{red}{red balls} do you expect the selected urn to have?}}
    \item \textbf{\textit{What is your uncertainty level about this percentage?}}
\end{enumerate}

\vspace{2mm}

\includegraphics[width=15cm]{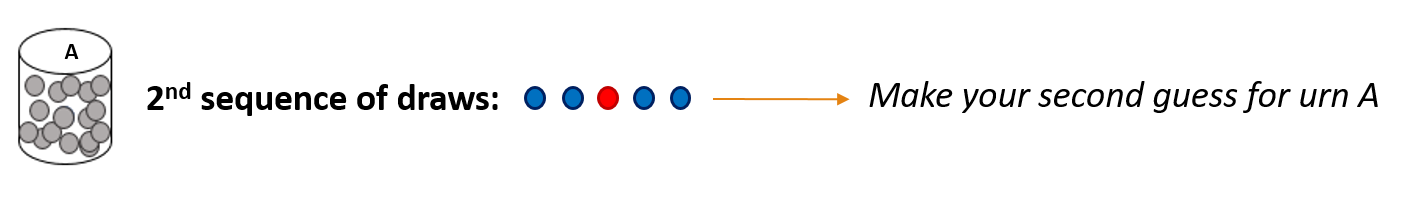}

\end{itemize}

In order to further help you with your guesses a dynamic graph of your choice will be provided. Please watch the following video (next screen) to understand how this works.

\vspace{2mm}

------------------------------------------[Page Break]---------------------------------------

\vspace{2mm}

\href{https://www.youtube.com/watch?v=g5Eg-aMp_5g&ab_channel=The5thBeatle2010}{Click here to see the explanatory video}

\vspace{2mm}

------------------------------------------[Page Break]---------------------------------------

\vspace{2mm}

{\large \textbf{Your Payment}}

\vspace{2mm}

You can earn up to \textbf{€29,70} in this experiment. In particular, your payment is broken down as follows:

\begin{itemize}
    \item You will receive €5 for taking the time to complete this experiment.
    \item You will receive up to €10 for your responses related to the guessing tasks.

How much of this amount (€10) you receive depends on the actual percentage of red balls in the selected urn. You can get money for \textbf{EVERY SINGLE ONE} of your guesses.

The payment rule we use, is optimized so that in order to \textbf{maximize your expected payoff}, you should \textbf{ALWAYS give your best estimate} of the percentage of red balls in the selected urn. In the same manner, the payment rule we use, is also optimized so that in order to \textbf{maximize your expected payoff}, you should \textbf{ALWAYS give your best estimate} of your uncertainty level.
 
\item You can receive up to €14,70 as \textbf{extra payment}.

This extra payment is divided across the 30 guessing tasks. Whether or not a guessing task has an extra payment attached depends on whether a \textcolor{red}{\textbf{dollar urn}} has come up. (See picture below. Urn A is in this example, a dollar urn). In particular, a dollar urn \textbf{will come up randomly} in 15 of the 30 guessing tasks.

\begin{figure}[ht]
    \centering
    \includegraphics[width=2cm]{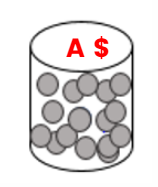}
    \label{fig:Dollar Urn}
\end{figure}

If a dollar urn comes up, you will receive \textcolor{red}{\textbf{as many cents as red balls}} the selected dollar urn has. For example, if the selected dollar urn has 50 red balls you will get 50 cents. 
\end{itemize}

At the end of the experiment, you will be informed about the number of red balls in each urn and your total payment.
If you want to know more about the details of the payment rule, you can let me know after the experiment or write an email to p.gonzalezfernandez@maastrichtuniversity.nl

------------------------------------------[Page Break]---------------------------------------


{\large \textbf{Get Familiar with the Tools}}

Before you answer the comprehension questions you have the chance to get familiar with the guesses, the payment and the sliders with a trial guessing task. 

------------------------------------------[Page Break]---------------------------------------

\begin{figure}[ht]
    \centering
    \begin{minipage}{0.48\textwidth}
        \centering
        \includegraphics[width=\linewidth]{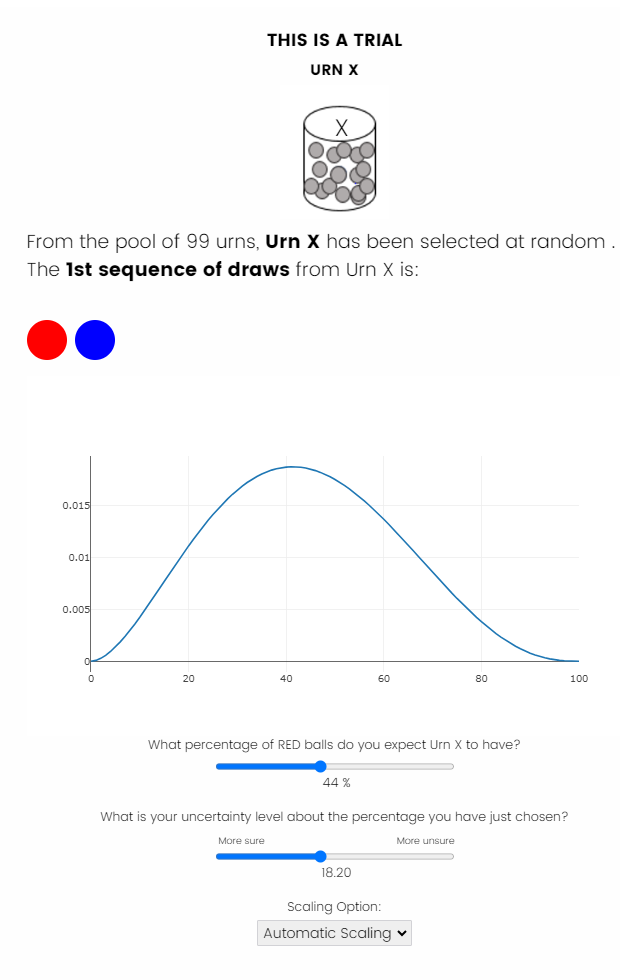}
    \end{minipage}
    \hfill
    \begin{minipage}{0.48\textwidth}
        \centering
        \includegraphics[width=\linewidth]{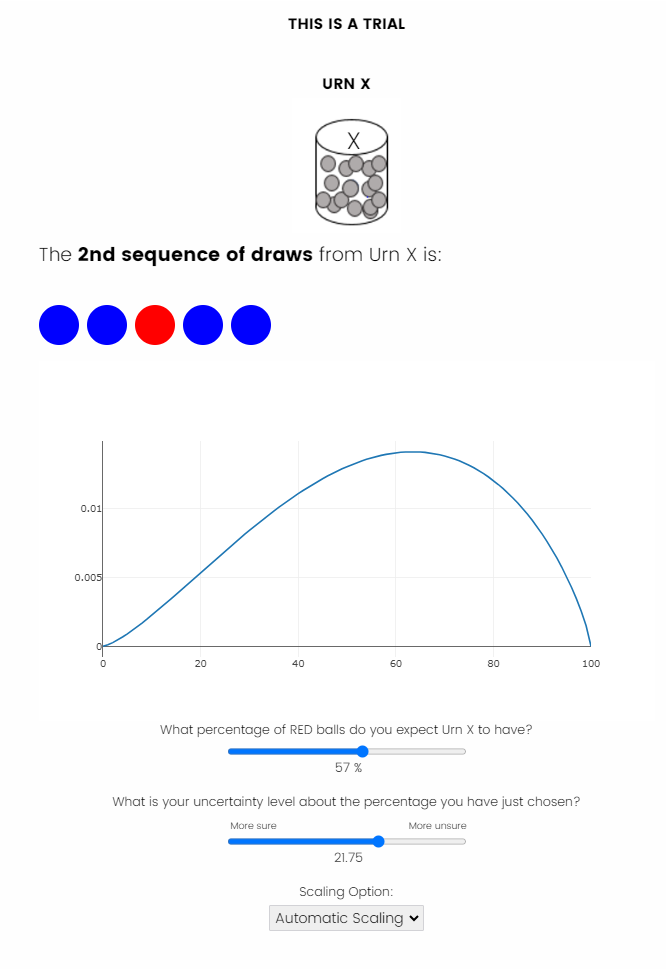}
    \end{minipage}
\end{figure}

\end{document}